\documentclass[twocolumn,times,tighten]{aastex631}
\usepackage{amsmath,amstext}
\usepackage[english]{babel}
\usepackage[utf8x]{inputenc}
\usepackage{textgreek}
\usepackage{graphicx}
\usepackage{xspace}

\newcommand{\teff}{$T_\mathrm{eff}$\xspace}
\newcommand{\logg}{$\log g$\xspace}
\newcommand{\feh}{[Fe/H]\xspace}
\newcommand{\microturb}{$\xi_\mathrm{micro}$\xspace}

\newcommand{\kms}{km\,s$^{-1}$\xspace}
\newcommand{\mic}{$\mu \mathrm m$\xspace}
\newcommand{\msun}{$\mathrm{M}_\odot$\xspace}
\newcommand{\loggf}{$\log gf$\xspace}

\begin{document}

\title{Rubidium Abundances in Cool Giants from High-Resolution H-band Spectra:\\ A New Diagnostic for Galactic Chemical Evolution}


\correspondingauthor{Nils Ryde}
\email{nils.ryde@fysik.lu.se}

\author[0000-0001-6294-3790]{Nils Ryde}
\affil{Division of Astrophysics, Department of Physics, Lund University, Box 118, SE-22100 Lund, Sweden}

\author[0009-0005-2049-3847]{Jess Kocher}
\affil{Division of Astrophysics, Department of Physics, Lund University, Box 118, SE-22100 Lund, Sweden}

\author[0000-0002-6077-2059]{Govind Nandakumar}
\affil{Aryabhatta Research Institute of Observational Sciences, Manora Peak, Nainital 263002, India}
\affil{Division of Astrophysics, Department of Physics, Lund University, Box 118, SE-22100 Lund, Sweden}

\author[0000-0001-9853-2555]{Henrik Hartman}
\affil{Materials Science and Applied Mathematics, Malm\"o University, SE-205 06 Malm\"o, Sweden}

\author[0000-0000-0000-0000]{Marta Molero}
\affil{Institut f\"ur Kernphysik, Technische Universit\"at Darmstadt, Schlossgartenstr. 2, Darmstadt D-64289, Germany}
\affil{INAF, Osservatorio Astronomico di Trieste, Via Tiepolo 11, I-34131 Trieste, Italy}

\author[0000-0002-4912-8609]{Henrik Jönsson}
\affil{Materials Science and Applied Mathematics, Malm\"o University, SE-205 06 Malm\"o, Sweden}

 \author[0000-0001-7875-6391]{Gregory Mace}  
 \affil{Department of Astronomy and McDonald Observatory,  University of Texas at Austin, 2515 Speedway, Stop C1400, Austin,
TX 78712-1205, USA}

 \author[0000-0002-8378-1062]{Erica Sawczynec}  
 \affil{Department of Astronomy, University of Texas at Austin,  2515 Speedway, Stop C1400, Austin, TX 78712-1205, USA}

 \author[0000-0001-6909-3856]{Kyle F. Kaplan}  
 \affil{Department of Astronomy, University of Texas at Austin,  2515 Speedway, Stop C1400, Austin, TX 78712-1205, USA}





\begin{abstract}
The Galactic Center and inner disk of the Milky Way contain complex stellar populations obscured by heavy dust extinction. To study their chemical composition, high-resolution near-infrared (near-IR) spectroscopy is necessary. Expanding the set of elements measurable in the near-IR, especially neutron-capture elements, improves our ability to trace nucleosynthesis and Galactic chemical evolution. This work aims to identify and characterize a spectral line suitable for determining rubidium (Rb) abundances. Rb is produced in roughly equal parts by the r- and s-processes. We analyze high-resolution ($R = 45,000$) IGRINS near-IR spectra of 40 M giants in the solar neighborhood, most observed with Gemini South. We perform spectral synthesis of the Rb\,\textsc{i} line at 15289.48\,\AA, using  new \loggf values and including an astrophysical calibration of the blending Fe\,\textsc{i}  lines. The resulting [Rb/Fe] ratios are compared to other neutron-capture elements and interpreted with chemical evolution models. We demonstrate that the used Rb line is a reliable abundance indicator in M giants and the coolest K giants, but becomes too weak at higher temperatures. [Rb/Fe] shows a decreasing trend with metallicity, mirroring that of ytterbium (Yb), another mixed r-/s-process element. Our results agree with optical studies, validating the use of this near-IR line. Comparisons with chemical evolution models confirm that both s- and r-process sources are needed to explain the Rb trend. This work adds Rb to the list of elements measurable in high-resolution H- and K-band spectra, enabling studies of one more neutron-capture element in dust-obscured regions like the Galactic Center and inner disk.
\end{abstract}

\keywords{stars: abundances, late-type -- Galaxy:evolution, disk -- infrared: stars}

\section{Introduction}
\label{sec:intro}
\vspace{-5pt}

Different chemical elements form on different timescales and the galactic chemical evolution of the elements depends on factors such as the star formation rate, the influence of gaseous flows and mixing processes, as well as stellar yields \citep[see, e.g.,][]{matteucci:12,prantzos:18,Matteucci:2021}. As a result, abundance ratio trends as a function of metallicities offer a tool to study the evolution and properties of stellar populations. Ideally, one would aim to measure as many elements as possible, spanning a broad range of nucleosynthetic origins. The number of independent chemical dimensions has been a subject of debate \citep[see, e.g.,][]{mead}, but, for instance, the abundances of slow neutron-capture (s-process) elements add a specific dimension that differs from that of the $\alpha$-elements \citep{manea:23}. Low-mass Asymptotic Giant Branch (AGB) stars are responsible for most of the production of the main s-process elements, such as barium (Ba), yttrium (Y), cerium (Ce), and neodymium (Nd). Owing to the lower masses of these AGB stars \citep[1.3–3\,\msun;][]{Grisoni:2020}, compared to the more massive progenitors of core-collapse supernovae (SNe Type II; SNeII) that dominate the production of $\alpha$-elements, s-process elements are formed on longer timescales. 

To study all stellar populations in the Milky Way, including the interesting and important dust-obscured populations (e.g., in the Galactic Center), elemental abundances must be determined from infrared spectra. A range of high-resolution near-infrared (near-IR) spectrometers is now available, capable of covering large portions of one or more near-IR photometric bands. These include the Apache Point Observatory Galactic Evolution Experiment \citep[APOGEE;][]{apogee}, GIANO \citep{giano:06,Origlia:2014}, the Near-InfraRed Planet Searcher \citep[NIRPS;][]{nirps,pasquini2018eso-485}, the Immersion GRating INfrared Spectrograph \citep[IGRINS;][]{Yuk:2010, Wang:2010, Gully:2012, Moon:2012, Park:2014, Jeong:2014}, and IGRINS-II \citep{IGRINS2}. Several new instruments are also under development, such as the Multi-Object Optical and Near-infrared Spectrograph \citep[MOONS;][]{MOONS2020}, the ArmazoNes high Dispersion Echelle Spectrograph \citep[ANDES;][]{Andes}, and the Habitable Worlds Observatory \citep[HWO;][]{HWO}.

With help of these instruments, approximately two dozen elements have been identified as measurable through infrared spectroscopy. For instance, APOGEE \citep{apogee_dr17} derives abundances for 18 elements from H-band spectra: the CNO elements (C, N, O); the odd-Z elements (Na, Al, K); the $\alpha$-elements (Mg, Si, S, Ca); the iron-peak elements (Ti, V, Cr, Mn, Fe, Co, Ni); and the s-process element Ce. Additional elements with weak or blended spectral lines, beyond the capability of APOGEE’s automated pipeline, such as P, the weak-s element Cu, and an additional s-process element, Nd, are provided in the BACCHUS Analysis of Weak Lines in APOGEE Spectra (BAWLAS) catalog \citep{Hayes:2022}.

Using IGRINS spectra, which have a higher resolution than APOGEE and cover both the H and K bands, \citet{nandakumar:22,Nandakumar:24_21elements}  determined abundances for 22 elements of giants in the solar neighborhood, expanding the set of measurable elements from near-IR spectra to include  F \citep[e.g.,][]{Nandakumar:2023b,Ryde:2020}, Sc, Zn, Y, Ba, and Yb.  While many of the elements have lines in both bands, F, Sc, Y, and Ba can only be determined from the K band. Phosphorus can be determined in the H band, particularly in K giants, although this becomes increasingly challenging in cooler stars \citep{nandakumar:22}. For cool giants,  IGRINS spectra allow the retrieval of 21 elements, while APOGEE provides abundances for 14 elements in similarly cool stars,  including the careful analysis of weak lines in the BAWLAS study.  These advancements in near-IR spectroscopy enable the chemical evolution of even dust-obscured regions of the Galaxy to be studied effectively, as demonstrated for the stellar populations in the very center of the Milky Way: The Nuclear Star Cluster \citep{ryde:25,NSC_all:25} and the Nuclear Stellar Disk (Ryde et al. 2025b). Several other studies of chemical abundances have been conducted based on high-resolution spectra observed with IGRINS, such as those for very metal-poor stars \cite[e.g.][]{afsar:16,aldo:20,aldo:25}, carbon stars \citep{garcia:23}, and evolved field stars, as well as those in open and globular clusters \cite[e.g.][]{Afsar:2018,bocek:19,bocek:20,montelius:22,brady:23,holanda:24,odzemir:25}.

In terms of the important neutron-capture elements, we thus see that also Y (s/r = 70/30)\footnote{These s-process elements are produced through a combination of the s- and r-processes. For Ce and Nd, the s-process dominates, with s/r ratios of 85/15 and 60/40, respectively, in the Solar system isotopic composition \citep{bisterzo:14,prantzos:20}.}, 
Ba \citep[s/r = 90/10;][]{Nandakumar:ba}, and Yb \citep[s/r = 40/60;][]{montelius:22} can be readily measured in high resolution IGRINS spectra, in addition to Cu, Ce, and Nd.  Abundances of Cu, Ce, Nd, and Yb can be obtained from H-band spectra, while the K-band provides access to Y, and Ba.

In this work, we add rubidium (Rb) to the list of elements measurable from high-resolution near-infrared spectra of cool giants. Rb is of particular interest because its cosmic production is approximately evenly split between the r- and s-processes \citep{sneden:08,prantzos:20}. This is similar to ytterbium, which is slightly more dominated by the r-process, with an s/r contribution ratio of 40/60. Here, we identify and characterize the near-infrared Rb\,\textsc{i} line at 15289.48\,\AA\ in high-resolution IGRINS spectra. Thus, Rb constitutes a new neutron-capture element accessible for abundance studies in the near-infrared.

 \begin{deluxetable*}{c c c c c c c}
 \tablecaption{Stellar parameters and the thin/thick-disk population membership for the stars in our sample, from \citet{Nandakumar:2023}.}
\tablehead{
\colhead{Name} & \colhead{T$_\mathrm{eff}$} & \colhead{$\log g$} & \colhead{[Fe/H]} & \colhead{$\xi_\mathrm{micro}$} & \colhead{[Rb/Fe]\tablenotemark{a}} & \colhead{Population\tablenotemark{b}}   \\
 \colhead{} & \colhead{$[\mathrm{K}]$} & \colhead{(cgs)} & \colhead{dex} & \colhead{[\kms]} & \colhead{}  &  
 } \label{table:parameters}
\startdata
2M05484106-0602007 & 3490  &  0.48  &  -0.28  &  2.03     & 0.01&   thin        \\
 2M06035110-7456029& 3562  &  0.48  &  -0.51  &  2.14   &   0.11&  thick    \\
 2M06035214-7255079& 3742  &  1.08  &  0.0  &  1.78     &   0.02&   thick    \\
 2M06223443-0443153& 3521  &  0.4  &  -0.52  &  2.19   &  0.2  &   thin   \\
 2M06231693-0530385& 3484  &  0.32  &  -0.55  &  2.09  &0.33&      thin   \\
 2M06520463-0047080& 3581  &  0.67  &  -0.21  &  2.15  &  0.0  &   thin   \\
 2M06551808-0148080& 3606  &  0.52  &  -0.56  &  1.96  &  0.26  &   thin   \\
 2M10430394-4605354& 3568  &  0.96  &  0.25  &  1.83   &  -0.12  &  thin   \\
 2M13403516-5040261& 3528  &  0.61  &  -0.15  &  1.92  &  -0.21  &  thin   \\
 2M14131192-4849280& 3504  &  0.61  &  -0.08  &  1.81  &  -0.07  &  thin   \\
 2M14240039-6252516& 3474  &  0.69  &  0.12  &  1.94   &  -0.09  &  thin   \\
 2M14241044-6218367& 3543  &  0.8  &  0.11  &  1.95    &  -0.33  &  thin   \\
 2M14260433-6219024& 3386  &  0.55  &  0.13  &  1.82   &  -0.13  &  thin   \\
 2M14261117-6240220& 3387  &  0.52  &  0.08  &  1.92   &  0.06  &  thin   \\
2M14275833-6147534& 3453  &  0.63  &  0.08  &  1.91   &  -0.08  &  thin   \\
2M14283733-6257279& 3465  &  0.62  &  0.04  &  1.83   &  0.01  &  thin   \\
2M14291063-6317181& 3430  &  0.54  &  0.0  &  1.95    &  0.01  &  thin   \\
2M14311520-6145468& 3499  &  0.62  &  -0.06  &  2.01  &  -0.1  &  thin   \\
2M14322072-6215506& 3639  &  0.89  &  -0.0  &  1.76   &  -0.47  &  thin   \\
2M14332169-6302108& 3524  &  0.56  &  -0.25  &  1.98  &  0.05  &  thin   \\
2M14332869-6211255& 3664  &  1.11  &  0.23  &  1.99   &  -0.18  &  thin   \\
2M14333081-6221450& 3430  &  0.55  &  0.02  &  1.92   & -0.06 &    thin   \\
2M14333688-6232028& 3425  &  0.54  &  0.02  &  1.87   &  -0.05  &  thin   \\
2M14345114-6225509& 3442  &  0.68  &  0.18  &  1.85   &  0.05  &  thin   \\
 2M14360935-6309399& 3446  &  0.61  &  0.08  &  1.99   &  0.02  &  thin   \\
 2M14371958-6251344& 3650  &  0.98  &  0.1  &  1.8     &  -0.15  &  thin   \\
 2M14375085-6237526& 3582  &  0.96  &  0.23  &  1.8    &  0.04  &  thin   \\
 2M15161949+0244516&  3691  &  0.76  &  -0.4  &  1.98  &   0.09  &  thick  \\
 2M17584888-2351011& 3564  &  0.95  &  0.25  &  2.2    &  -0.10  &  thin   \\
 2M18103303-1626220& 3347  &  0.46  &  0.09  &  1.98   &  0.12  &  thin   \\
 2M18142346-2136410& 3390  &  0.48  &  0.01  &  1.96   &  0.12  &  thin   \\
 2M18191551-1726223& 3434  &  0.59  &  0.07  &  1.93   &  -0.04  &  thin   \\
 2M18522108-3022143&  3578  &  0.45  &  -0.59  &  2.26 &  0.23   &  thick  \\
 KIC10649021 &  3861 & 1.10    &  -0.31    & 1.68  & 0.07  &  thin \\
 HD132813&  3457  &  0.43  &  -0.27  &  1.88 &   0.19  & thin   \\
 HD175588&  3484  &  0.49  &  -0.04  &  2.24 &  -0.04  & thin   \\
 HD89758&  3807  &  1.15  &  -0.09  &  1.65 &  -0.19  & thin   \\
 HD224935&  3529  &  0.64  &  -0.10  &  2.01 &  -0.09  & thin   \\
 HD101153&  3438  &  0.51  &  -0.07  &  2.03 &  -0.04  & thin   \\
 HIP54396&  3459  &  0.50  &  -0.15  &  1.86 &   0.15  &    thin   \\
\enddata
\tablenotetext{a}{We normalize to the solar value $\log \epsilon_\odot(\mathrm{Rb}) = 2.60 \pm 0.10$  \citep{Grevesse:1998}.}
\tablenotetext{b}{Stellar population assignment, either thin or thick disk, see \citet{Nandakumar:2023}.}
\end{deluxetable*}

\section{Observations and Stellar Parameters}
\label{sec:observations} 
We have analyzed the rubidium abundance in 40 M giants ($3350<$\teff$<3900$ K), see Table\,\ref{table:parameters}. The observations of and parameter determination for the stars were presented in \citet{Nandakumar:2023} and later also used in \citet{Nandakumar:2023b,Nandakumar:24_21elements,Nandakumar:ba}. To summarize the observations, we obtained 33 high-resolution, near-infrared spectra using IGRINS, mounted on the Gemini South telescope \citep{Mace:2018}, in service mode during January–April 2021. An additional six spectra were taken from 
the Raw and Reduced IGRINS Spectral Archive 
\citep[RRISA;][]{rrisa,rrisa:25}, comprising stars observed at McDonald Observatory \citep{Mace:2016}. 

The spectra cover the full H and K bands ($1.45-2.5$\,\mic) at a resolving power of $R \sim 45{,}000$, and with an average signal-to-noise ratio per resolution element generally well above 100 \citep[see ][]{Nandakumar:2023}.  The spectra were extracted from the observed data by using the standard IGRINS Pipeline Package \citep[IGRINS PLP;][]{kaplan_plp} and the telluric lines were reduced using early-type standard stars observed at matching airmass. Orders were normalized and stitched using {\tt iraf} \citep{IRAF} routines, excluding low-S/N edges. Finally, the spectra were corrected to the laboratory rest frame. To take any modulations in the continuum levels of the spectra into account, we carefully defined the continuum in the segment (a 30\,\AA\ window) where the Rb\,\textsc{i}  line lies for every star.

The stellar parameters\footnote{The effective temperature (\teff), surface gravity (\logg), metallicity (\feh), and microturbulence (\microturb).} of the stars are provided in Table \ref{table:parameters} and were derived in \citet{Nandakumar:2023} using the spectral synthesis code Spectroscopy Made Easy \citep[SME;][]{sme,sme_code} in combination with MARCS 1D spherical model atmospheres \citep{marcs:08}. An iterative approach was adopted in which effective temperature ($T_{\rm eff}$), surface gravity ($\log g$), metallicity $[\mathrm{Fe/H}]$, microturbulence ($\xi_{\rm micro}$), and C and N abundances were simultaneously optimized to fit a selected set of atomic and molecular lines (Fe, CO, CN, and OH). The OH lines, in particular, were used to constrain $T_{\rm eff}$, assuming a fixed [O/Fe] abundance based on the thin- or thick-disk chemical trends adopted from \citet{Amarsi:2019}, see \citet{Nandakumar:2023}.
The surface gravity, \logg, was found by interpolating in Yonsei-Yale isochrones \citep{Demarque:2004} assuming stellar ages of 3–10 Gyr. Typical uncertainties are estimated to be $\pm100$\,K in $T_{\rm eff}$, $\pm0.2$\,dex in \logg, $\pm0.1$\,dex in \feh, and $\pm0.1$\,\kms\ in $\xi_{\rm micro}$, see also \citet{Nandakumar:2023}. 

In Table \ref{table:parameters}, we also indicate the stellar population each star is ascribed to, either the high-alpha sequence (thick disk) or the low-alpha sequence (thin disk); see \citet{Nandakumar:2023}.

\section{Analysis}
\label{sec:analysis} 


We have successfully used the Rb $\lambda15289$ line at 15289.48\,\AA\ to measure the Rb abundance in 40 M giants using high-resolution IGRINS spectra at $R\sim 45000$.  Here, we discuss the near-IR Rb\,\textsc{i} line and the approach used to model it.

\begin{deluxetable*}{l c c c c c c}
 \tablecaption{Atomic data for  Rb\,\textsc{i} near-IR lines. }
\tablehead{
\colhead{Element} & \colhead{$\lambda_{\mathrm{air}}$ [\AA]} & \colhead{$\log gf$} & \colhead{$\log gf$\tablenotemark{a}} & \colhead{$E_\mathrm{low}$ [eV] } & \colhead{$E_\mathrm{up}$ [eV]} & \colhead{Transition} \\
 & \colhead{\citet{Johansson:61}} & \colhead{\citet{K14}} & \colhead{\citet{M:16}} & \colhead{\citet{sansonetti:06}} & \colhead{\citet{sansonetti:06}  }
 }   \label{tab:rb_lines}
\startdata
Rb\,\textsc{i} & 14752.415 & {\bf $+0.167$}   & +0.126 & 1.5596 & 2.3998 & $4p^6\,5p\,\,^2\!P^\circ_{1/2} \rightarrow 4p^6\,4d\,\,^2\!D_{3/2}$ \\
Rb\,\textsc{i} & 15288.437 & {\bf $-0.535$} & $-0.580$ & 1.5890 & 2.3998 & $4p^6\,5p\,\,^2\!P^\circ_{3/2} \rightarrow 4p^6\,4d\,\,^2\!D_{3/2}$ \\
Rb\,\textsc{i} & 15289.480 & {\bf $+0.420$}  & +0.374 & 1.5890 & 2.3997 & $4p^6\,5p\,\,^2\!P^\circ_{3/2} \rightarrow 4p^6\,4d\,\,^2\!D_{5/2}$ \\
\enddata
\tablenotetext{a}{\citet{M:16} provides the absorption oscillator strengths of the transitions, $f_\mathrm{lu}$, from which the  \loggf = $\log [(2J+1)\cdot f_\mathrm{lu}$] is derived}
\end{deluxetable*}


\begin{figure}
    \centering
    \includegraphics[trim=0cm 0cm 0cm 0cm, clip, width=1.0\linewidth]{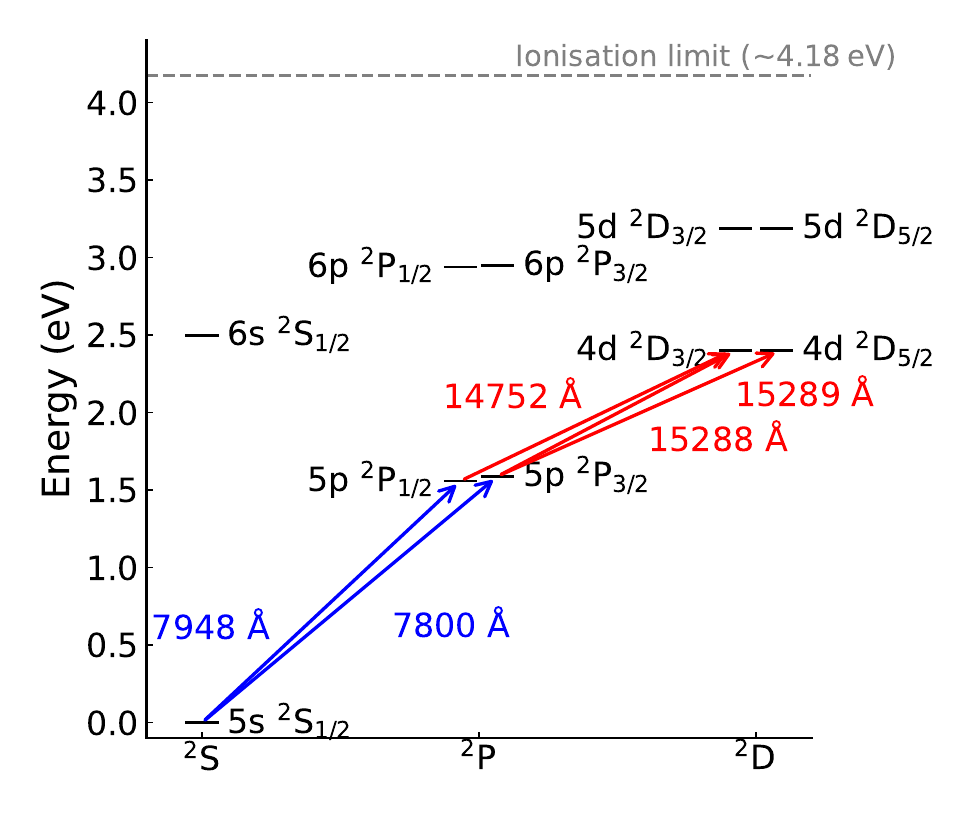}
    \caption{Partial term diagram for Rb\,\textsc{i} showing the first energy levels of the lowers terms. The two resonance lines are marked in blue and the three lines in the H-band multiplet are marked in red. It is the $\lambda15289$ ($5p\,\,^2\!P^\circ_{3/2} \rightarrow 4d\,\,^2\!D_{5/2}$) transition that is analyzed here. }
    \label{fig:term}
\end{figure}

\subsection{The Rb\,\textsc{i} Line at $\lambda15289$ }

Rubidium is an alkali metal (Z=37), located just below potassium and sodium in the periodic table, and just to the left of the light first peak s-elements Sr and Y. Like Na and K, it has a single valence electron that is easily ionized. 
Rubidium has the electron configuration [Kr]5s$^1$ and a very low ionization potential of 4.2\,eV \citep{sansonetti:06}, which is lower than that of both Na and K. The spectrum of Rb thus resembles that of an effective one-electron system outside closed shells, making it relatively simple, provided that high excitation levels are avoided. The Rb\,\textsc{i} line studied here, $\lambda15289$, is part of the $5p\,^2P \rightarrow 4d\,^2D$ multiplet (see Figure\,\ref{fig:term}) that consists of three lines with different $J$-values: $1/2\rightarrow 3/2$, $3/2\rightarrow 3/2$, and $3/2\rightarrow 5/2$. 
The lines in the multiplet lie at $14752.415$\,\AA\, \citep{M:16}, 15288.437\,\AA, and 15289.480\,\AA. Although lying within the IGRINS range, the first line at $14752.415$\,\AA\ is at the edge of the H-band which is heavily affected by telluric lines. The wavelength region is also heavily blended and Rb only makes up a fraction of the feature at this wavelength. In the M giants analyzed here, the line strength of the weaker of the two remaining Rb\,\textsc{i} lines, 15288.437\,\AA, is only one-fifth  of that of the line at 15289.480\,\AA.  The latter line was also discussed by \citet{Smith:2021} in the context of APOGEE; however, it has not been detected in APOGEE spectra due to the survey’s lower spectral resolution. The atomic data for all three lines in the $5p\,^2P \rightarrow 4d\,^2D$ transition are listed in Table~\ref{tab:rb_lines}.

Just as Na has its well-known resonance doublet at 5890 and 5896\,\AA, Rb exhibits a resonance doublet at 7800.2 and 7947.6\,\AA. The lower level of the Rb doublet at $\lambda15289$, studied here, corresponds to the upper level of the resonance line at $\lambda7800$, see Figure\,\ref{fig:term}. The naturally occurring isotopes are $^{85}$Rb and $^{87}$Rb, with $^{85}$Rb making up approximately 72\% of the natural abundance. It is dominated by the r-process \citep[r/s$\sim$70/30; ][]{prantzos:20}, while the heavier isotope stems almost exclusively from the s-process. Both isotopes have non-zero spin, allowing for hyperfine structure splitting of the levels and thus the lines. However, the largest splitting is expected from levels with an unpaired s-electron and the current transitions have p- and d-orbitals where the hyperfine splitting in general is much smaller \citep{spectrophysics}. We thus do not expect a large hyperfine structure splitting. In addition, since the lines are weak in the spectrum, the hyperfine structure de-saturation effect is very small, just like any line broadening effect.

\subsection{Modelling the Rb\,\textsc{i} Line}
  The Rb abundances are determined by fitting  a synthetic spectrum to the observed one. The synthetic spectra were computed using the Python version of the Spectroscopy Made Easy \citep[PySME;][]{sme,sme_code,pysme:23} tool, which performs spherical radiative transfer through stellar atmosphere models defined by the adopted fundamental parameters. The atmosphere models were selected by interpolating within the grid of one-dimensional MARCS models \citep{marcs:08}.

From the VALD database, only two of the lines in the multiplet have \loggf values, which are taken from \citet{K14} and originate from calculations by \citet{1968MNRAS.139..115W}. No experimental data exist, to our knowledge, for the infrared transitions. We have used more recent theoretical values from \citet{M:16}, which include experiment-tuned oscillator strengths $f_{lu}$-values, which are 0.669, 0.0657, and 0.592, respectively. Based on these, we have derived new \loggf values for all three lines in the multiplet, which are presented in column 4 in Table\,\ref{tab:rb_lines}. These \loggf from \citet{M:16} are around 0.05 dex smaller then the data from \citet{1968MNRAS.139..115W,K14}.

\begin{figure}
    \centering
    \includegraphics[trim=1.95cm 0cm 3cm 1.5cm, clip, width=\linewidth]{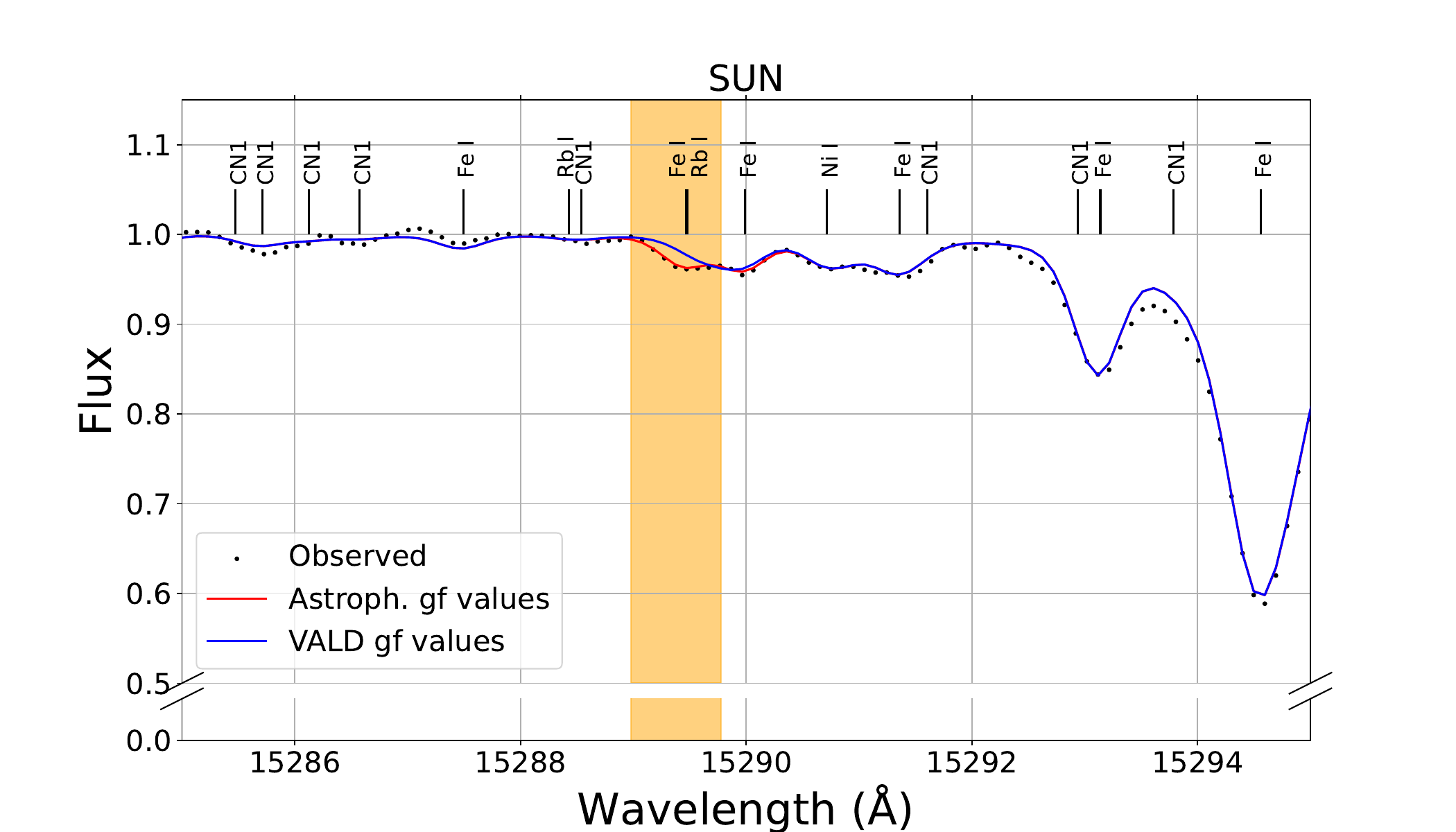}
    \caption{Solar spectrum (black) observed as a flux spectrum off the asteroid Ceres, and synthetic spectra using line data from \citet{K14} (blue) versus using our adjusted \loggf values (red).}
    \label{fig:sunoriginal}
\end{figure}

\begin{figure}
\centering
        \includegraphics[trim=1.95cm 0cm 3cm 1.5cm, clip, width=\linewidth]{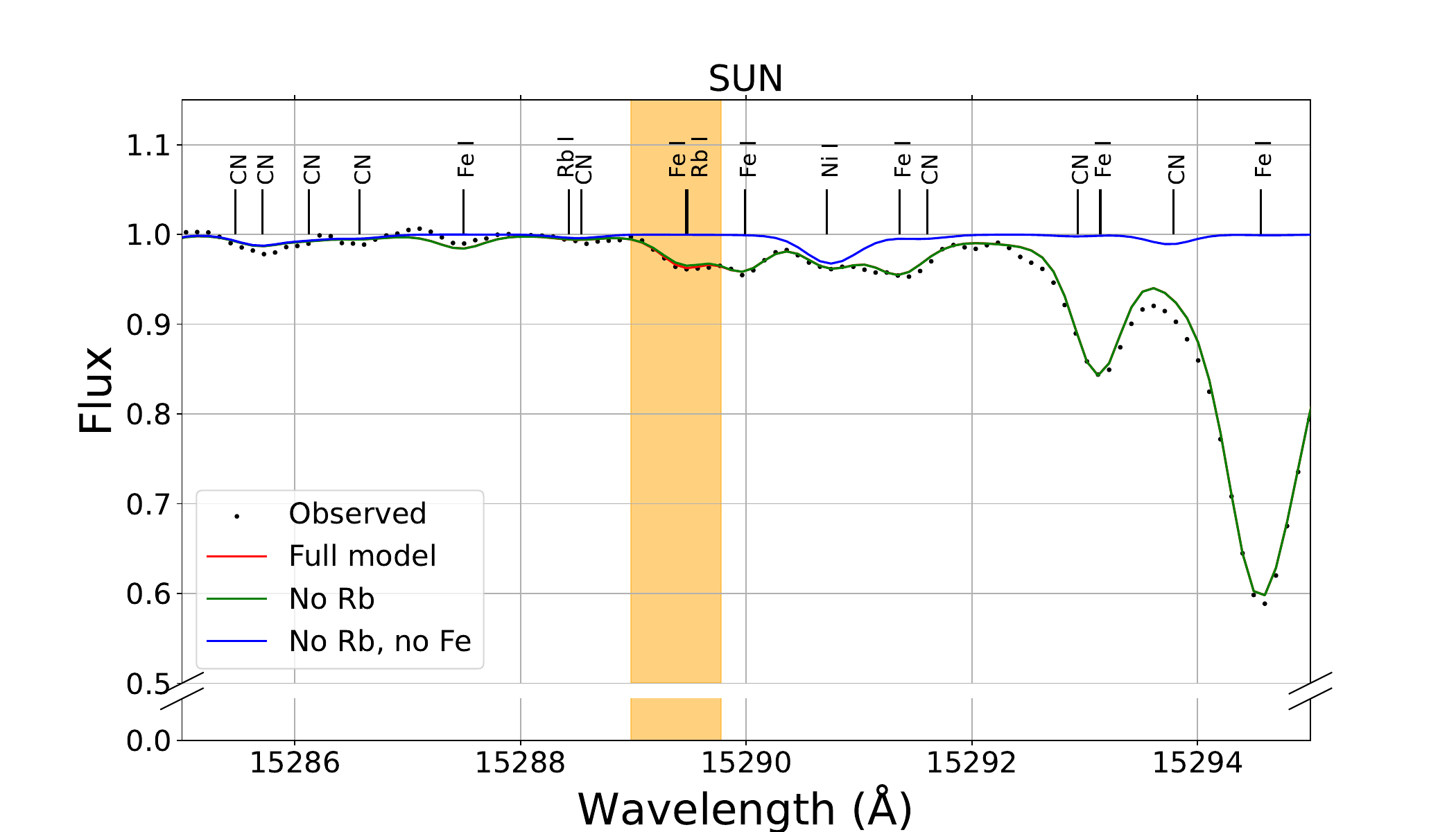}
        \caption{Observed (black dots) and synthetic (red line) spectra for the Sun (\teff $=5777$ K, \feh $=0.0$ dex). The synthetic spectrum shown in green omits the Rb\,{\sc i} line, only modeling blends; since it (approximately) overlaps with the red (fully modeled) spectrum, we can assume the Rb line to be insignificant in the Sun. The blue spectrum further omits any Fe\,{\sc i} lines, confirming that iron is by far the most prominent of the blends. The green and red spectra are modeled using our astrophysical \loggf\ values for the Fe\,{\sc i} lines, as listed in Table \ref{tab:NISTFeLines}.}
        \label{fig:sunblends}
\end{figure}

An initial major source of uncertainty in our rubidium abundance analysis was the modeling of spectral blends. The most problematic feature was the Fe\,{\sc i} line at $15 289.47$\,\AA, located just 0.01\,\AA\ from the center of the Rb\,\textsc{i} line. This feature is listed by \citet{1994Nave} as an "unidentified line due to iron" (see their Table 5). The corresponding energy levels were later identified by \citet{K14}, who predicted the line to be very weak (\loggf\ = $-5.176$). However, as shown in Figure~\ref{fig:sunoriginal}, this significantly underestimates the line strength observed in the solar spectrum. With a solar Rb abundance, the Rb line is very weak, and its contribution to the spectral feature is negligible in the Sun (see Figure\,\ref{fig:Rb_eqw}), allowing us to derive an astrophysical \loggf\ value for this Fe\,\textsc{i} line, independent of the uncertainties in the Rb line strength or the solar abundance of Rb. We also adjusted two other nearby iron lines: we removed the line at $15289.653$\,\AA\ and slightly increased the strength of the line at $15289.987$\,\AA. The updated spectral fit is shown in red in Figure~\ref{fig:sunblends}, and the adjusted atomic data are listed in Table~\ref{tab:NISTFeLines}, alongside the original values from the NIST database \citep{NIST_ASD,K14}. We chose to adopt the measured wavelength of the $15289.468$\,\AA\ line from \citet{1994Nave}, adjusting the \loggf\ value from \citet{K14} accordingly. Alternatively, we could have shifted the wavelength of the nearby line identified in \citet{K14}, which has a more plausible \loggf\ value. However, the correct identification remains uncertain. Either approach affects only the assigned excitation energy (6.2750\,eV vs. 6.2829\,eV), a difference of about 0.1\%, corresponding to a level population change of roughly 3\%. The effect of this difference in the relative behavior of the Fe lines for different stellar temperatures is negligible. The blending Fe lines will thus be treated appropriately in all our stars when determining the Rb abundances.

\begin{deluxetable*}{c  c c  c c c  c}
\tablecaption{Blending lines within 1\,\AA\ around the Rb line at 15289.48\,\AA\ from VALD \citep[][query date 2025-05-06]{vald,vald3}  and NIST \citep{NIST_ASD}.} \label{tab:NISTFeLines}
\tablehead{
\colhead{Element} & \colhead{$\lambda_{\mathrm{air}}$ [\AA]} & \colhead{$E_\mathrm{low}$ [eV]} & \colhead{$\lambda_{\mathrm{air}}$ [\AA]} & \colhead{$E_\mathrm{low}$ [eV]} & \colhead{\loggf} & \colhead{Astrophysical \loggf} \\
 & \colhead{\citet{1994Nave}} &  \colhead{\citet{1994Nave}} & \colhead{\citet{K14}} & \colhead{\citet{K14}} & \colhead{\citet{K14}}  }
\startdata
Fe\,\textsc{i}  & 15 289.468 & unknown & 15 289.466 &  6.283 & $-5.176$& $-0.876$\\
Fe\,\textsc{i}  & -- & -- & 15 289.653 & 6.275 & $-1.036$ & -- \\
Fe\,\textsc{i}  & 15 289.991 & 6.275 & 15 289.987 & 6.275&$-0.837$ & 
        $-0.737$\\
\enddata
\end{deluxetable*}

\begin{figure}
        \includegraphics[trim=1.95cm 0cm 3cm 1.5cm, clip, width=\linewidth]{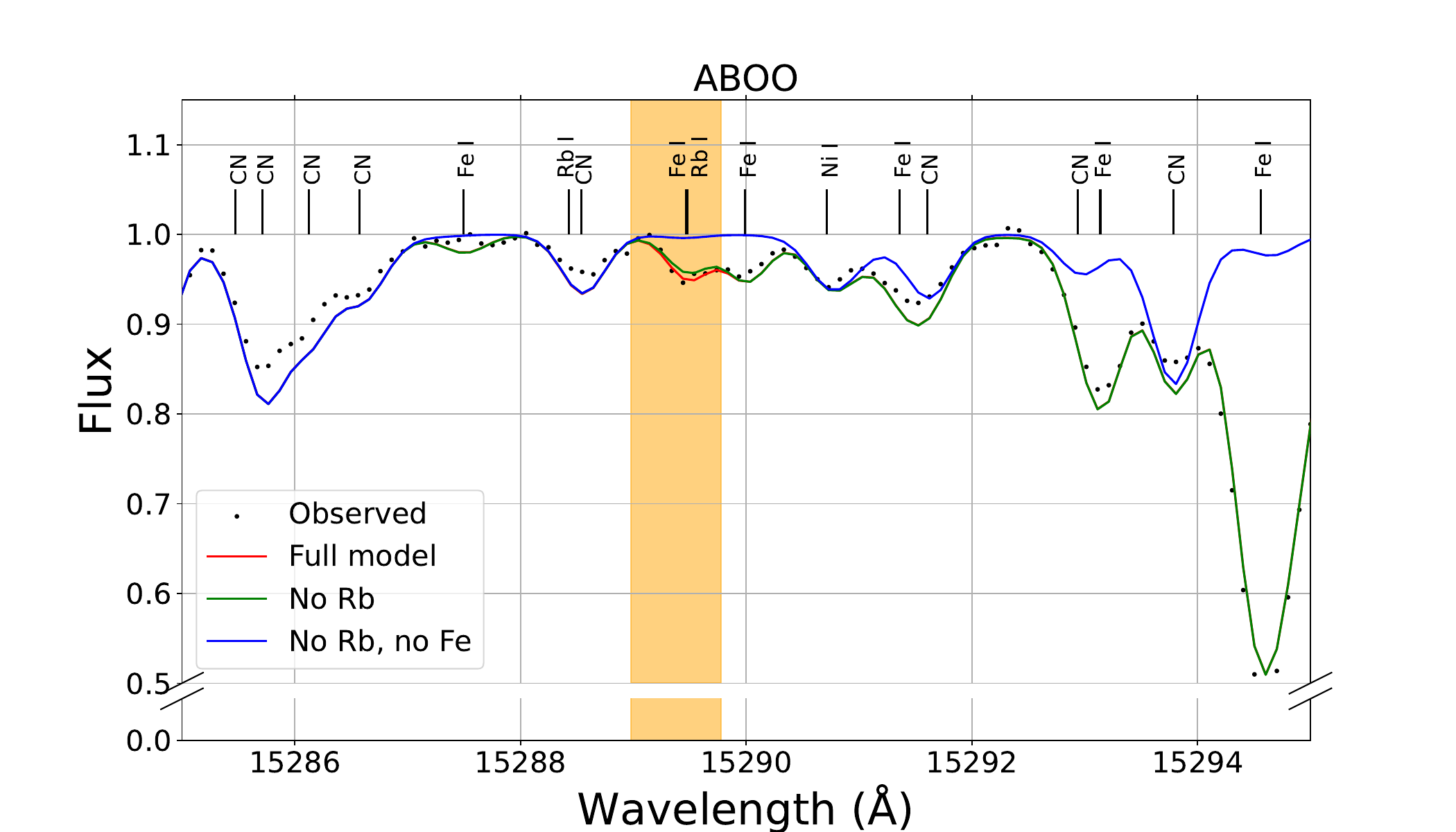}
        \caption{Observed (black dots) and synthetic (red line) spectra for Arcturus (\teff $=4308$ K, \feh $=-0.55$ dex). The green synthetic spectrum excludes the Rb\,{\sc i}  line and models only blends; its overlap with the full (red) model suggests Rb is negligible in Arcturus. The blue spectrum omits Fe\,{\sc i}  lines.}
        \label{fig:abooblends}
    \end{figure}

\begin{figure}
        \centering
        \includegraphics[trim=1.95cm 0cm 3cm 1.5cm, clip, width=\linewidth]{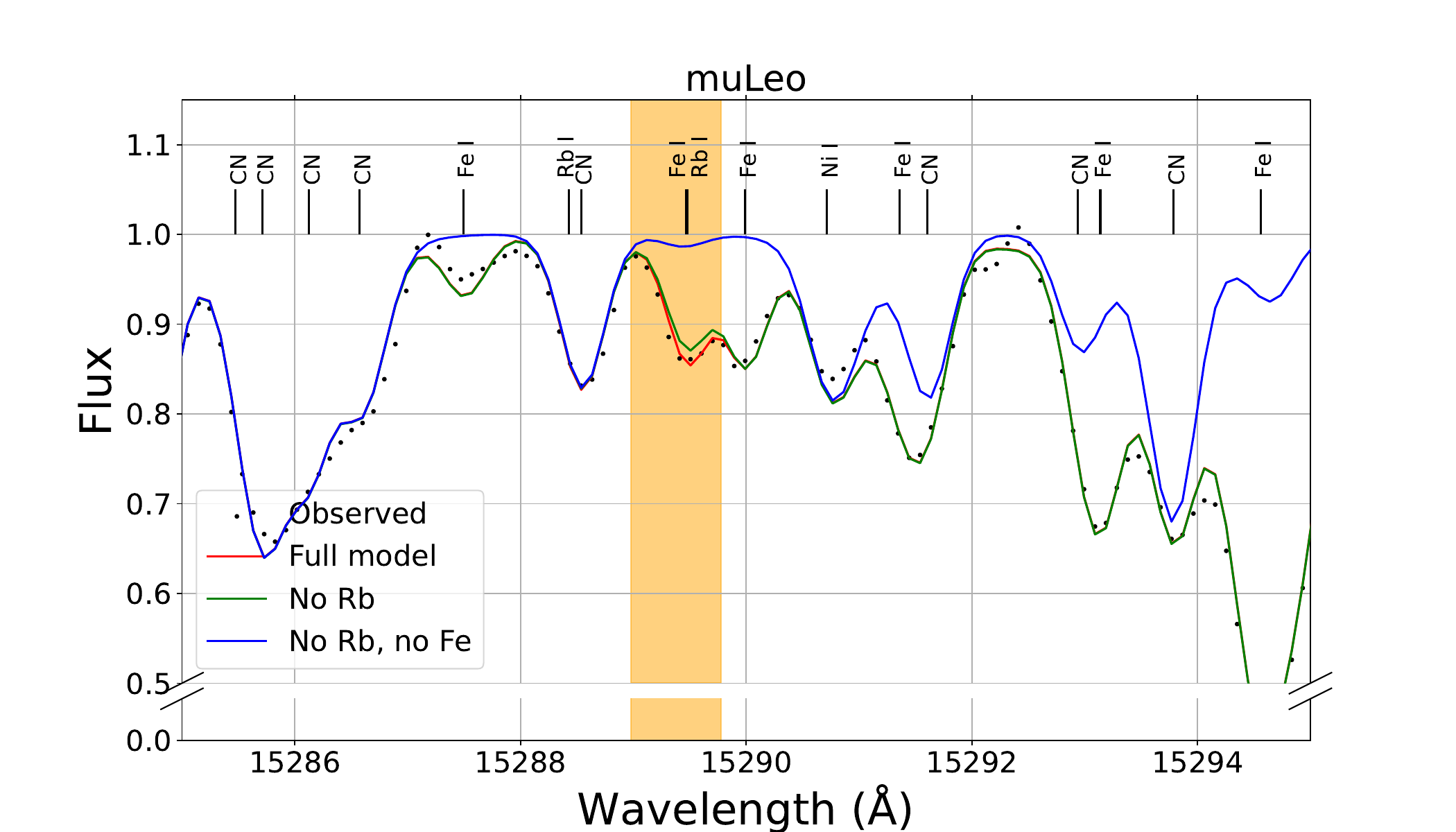}
        \caption{Observed (black dots) and synthetic (red line) spectra for $\mu$ Leo (\teff $=4494$ K, \feh $=0.27$ dex).  The green spectrum excludes Rb\,{\sc i}  and models only blends; The blue spectrum omits Fe\,{\sc i}   lines.}
        \label{fig:muleoblends}
    \end{figure}

To verify our adjustments, we considered a number of K giants with parameters determined from optical spectra by Jönsson et al. (in prep). Due to the very low ionization energy of Rb, the Rb line is essentially non-existent in stars with \teff $\gtrsim 4000$ K, making such stars a good sample to perform this test on. This is illustrated in Figures \ref{fig:sunblends}-\ref{fig:muleoblends}, where the synthetic spectra with and without the rubidium line (shown in red and green, respectively) match almost exactly. The Figure further shows that iron is essentially the only species we need to consider for the blends, as any remaining blends (shown in blue) are indistinguishable from the continuum in this region. With our adjusted iron lines, the region is modeled well for a range of temperatures and metallicities (Figures \ref{fig:sunblends}-\ref{fig:muleoblends}), thus confirming both the involved energy levels and oscillator strengths.

\begin{figure}
    \centering
    \includegraphics[trim=1.95cm 0cm 3cm 1.5cm, clip, width=\linewidth]{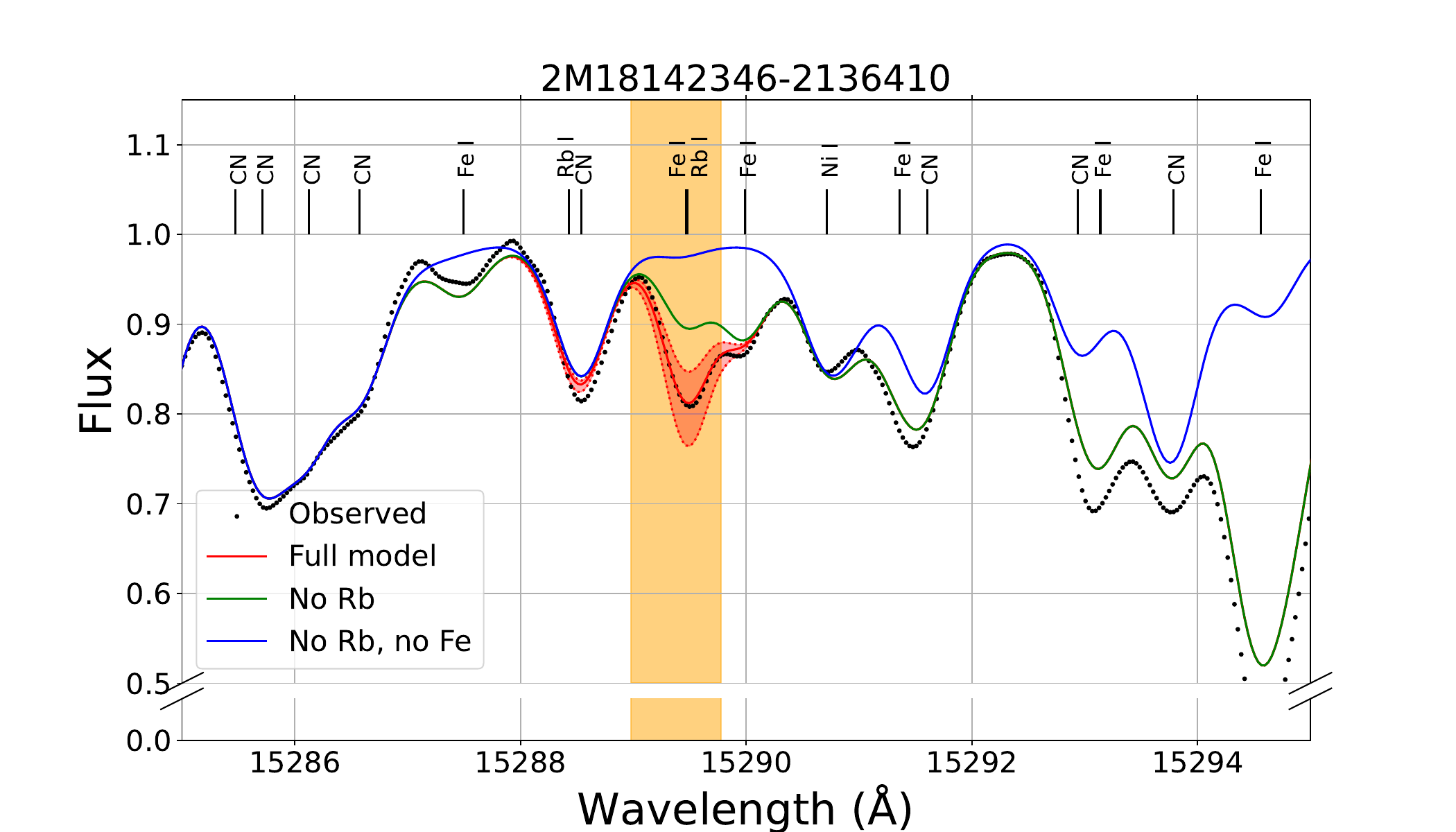}
    \caption{Same as Figure \ref{fig:sunblends} but for the M giant 2M18142346-2136410 (\teff $=3390$ K, \feh $=0.01$ dex, [Rb/Fe] = 0.07 dex). The difference between the models with (red) and without (green) rubidium shows that the Rb\,{\sc i} line is measurable for this star. The red shaded area marks the synthetic spectrum's sensitivity to $\pm 0.3$ dex in [Rb/Fe].}
    \label{fig:2M18142346-2136410Rb}
\end{figure}

In Figures \ref{fig:sunblends}-\ref{fig:muleoblends}, we showed that the Rb\,{\sc i} line is not measurable for warmer stars, and that Fe\,{\sc i} is the most relevant blend in this region. In Figure \ref{fig:2M18142346-2136410Rb} we now show the same kinds of model for the M giant 2M18142346-2136410 (\teff=3390\,K, \feh=0.01, [Rb/Fe]=0.07). The blue line (spectrum modeled without Rb or Fe) again confirms that there are no other blends we need to consider in detail, but this time we see a large difference between the red (fully modeled) and green (omitting Rb\,{\sc i} lines) spectra, implying that the Rb\,{\sc i} line is indeed measurable for this type of star. The red shaded area shows that a difference in rubidium abundance of $\pm 0.3$ dex would be clearly noticeable in the spectrum.

In the following, we report our Rb abundances in terms of  [Rb/Fe]. We adopt the solar reference value  $\log \epsilon_\odot(\mathrm{Rb}) = 2.60 \pm 0.10$ from \citet{Grevesse:1998}. For comparison, \citet{Asplund:2009} report a slightly lower value of $2.52\pm 0.1$, \citet{prantzos:20} report $2.45$, and \citet{abia:21} report $2.35\pm0.05$. The meteoritic abundance is given as $2.41\pm0.02$ by \citet{Grevesse:1998}, and $2.36\pm 0.03$ by \citet{Asplund:2009}.

\section{Results}
\label{sec:results}
\begin{figure}
  \includegraphics[trim=0cm 0cm 0cm 0cm, clip, width=\linewidth]{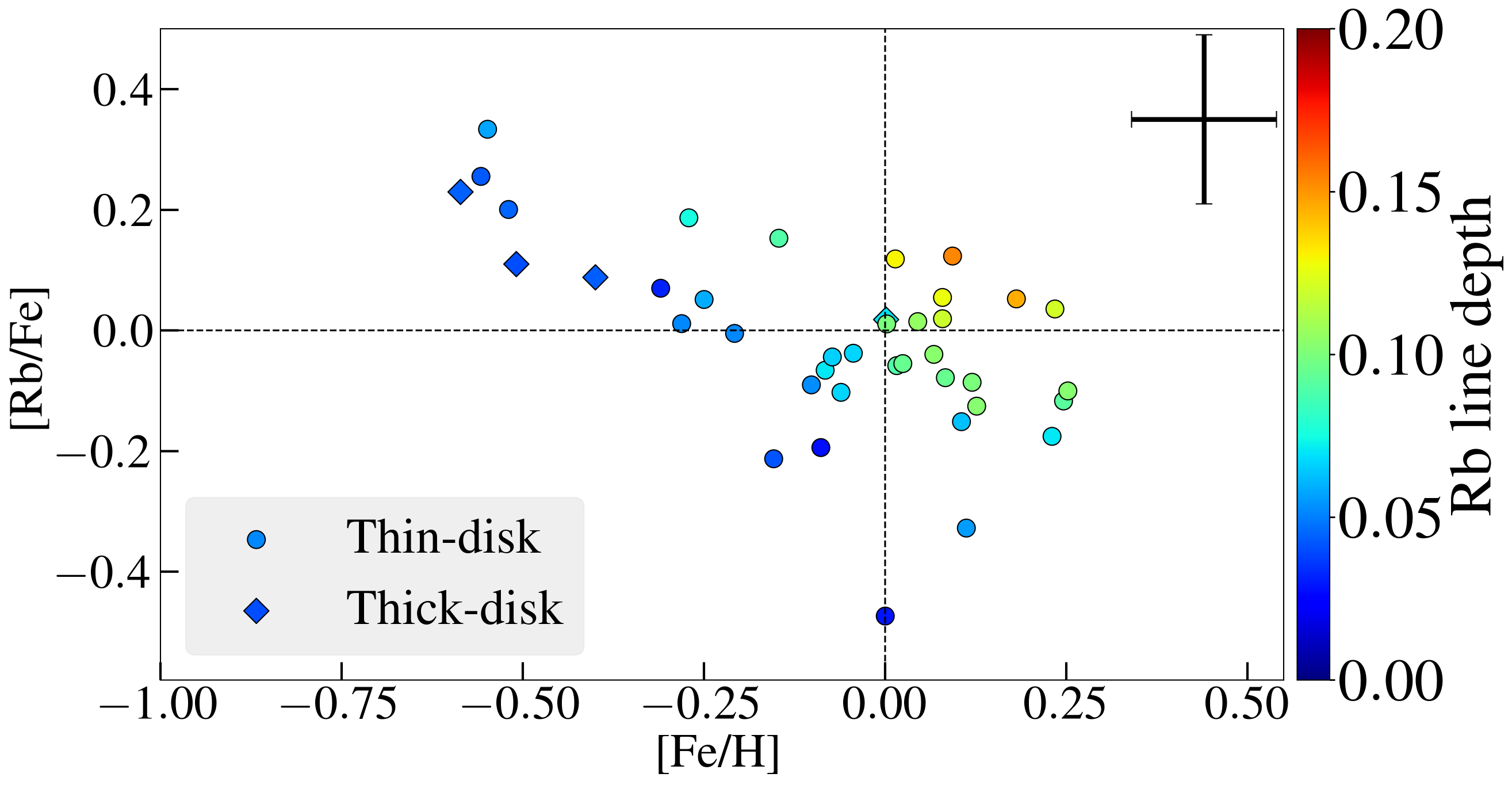}
  \caption{[Rb/Fe] versus [Fe/H] determined from the Rb\,\textsc{i} line at $\lambda15289$. The blue-green-red color-coding indicates the  strength of the line, with the blue ones being the weakest and red the strongest. The diamonds mark the thick-disk stars.}
  \label{fig:Rb_trend}%
\end{figure}

Our derived [Rb/Fe] abundance ratios are given in Table\,\ref{table:parameters}. Figure\,\ref{fig:Rb_trend} shows these data as the  rubidium abundance trend as a function of metallicity for the 40 M giants in our sample. We see a general downward trend with metallicity, similar to that of Yb \citep{montelius:22}, see Section\,\ref{sec:discussion}. The mean trend goes through the solar value.

In the plot, we specifically indicate the stars classified as thick-disk (high $\alpha$), based on the classifications in Table~\ref{table:parameters}. More stars are needed to draw any firm conclusions about the trend of the thick-disk stars compared to the thin-disk stars for subsolar metallicities. 

Uncertainties in the derived abundances arise from several sources and are challenging to quantify. Spurious features in the spectra, resulting from residuals left by telluric correction or low signal-to-noise levels, can distort spectral lines, preventing them from appearing as the clean Gaussian profiles typically expected. These irregularities may also introduce continuum undulations. In addition, broad absorption features caused by diffuse interstellar bands (DIBs; see, e.g., \citealt{dib:geballe1}) could further complicate the analysis. These features are likely caused by large molecules in the interstellar medium along the line of sight and typically correlate with reddening. DIBs have been identified in both optical and near-infrared spectra \citep[e.g.,][]{dib:geballe_nature,dib:cox,dib_crires}. A well-known DIB in the range 15268–15274\,\AA\ has been reported in several studies, including those based on IGRINS spectra \citep{dib:galazut} and APOGEE data \citep{dib:elya}. However, the Rb line lies 15\,\AA\ (corresponding to 300\,\kms) redward of this DIB, meaning it is unlikely to be affected. 

To highlight where the uncertainties in the derived Rb abundances are expected to be high, we have color-coded the rubidium abundances in Figure\,\ref{fig:Rb_trend} according to the strength of the Rb\,\textsc{i} line. We observe that the lowest Rb abundances are associated with the weakest lines, reducing the significance of these measurements.  

Additional uncertainties also arise from errors in the adopted stellar parameters. To estimate these uncertainties in the rubidium abundances arising from typical uncertainties\footnote{Typical uncertainties are $\pm$100 K in \teff, $\pm$0.2 dex in \logg, $\pm$0.1 dex in \feh, and $\pm$0.1 \kms\ in $\xi_\mathrm{micro}$, see Sect. \ref{sec:analysis}.} in the stellar parameters, we followed the method described in \citet{Nandakumar:2023} and \citet{Nandakumar:24_21elements}. For this purpose, we selected the metal-rich star 2M18142346-2136410 and recalculated the rubidium abundance 100 times, each time using stellar parameters drawn randomly from normal distributions centered on the measured values, with the standard deviations set to the typical uncertainties. The uncertainty in the rubidium abundance A(Rb) was then taken as half the range between the 84$^\mathrm{th}$ and 16$^\mathrm{th}$ percentiles of the resulting abundance distribution, yielding an estimate of $\pm$0.08 dex. In converting to the scale of [Rb/Fe], we further accounted for the 0.1 dex uncertainty of our metallicity values. We estimated an additional uncertainty of approximately 0.05 dex due to potential errors in the placement of continuum points around the rubidium line. Adding all sources of uncertainty in quadrature, we obtain a total typical uncertainty of 0.14 dex, which is shown as the error bar in Figure~\ref{fig:Rb_trend}. However, as noted earlier, and indicated in the Figure, the weaker the line, the larger the uncertainty in the derived abundance, given the signal-to-noise ratio.

\citet{korotin:20} investigated the non-LTE effects on the resonance lines of Rb\,{\textsc i} in cool stars. They derived a solar non-LTE abundance of $\log \epsilon_\odot(\mathrm{Rb}) = 2.35 \pm 0.05$, with non-LTE corrections of $-0.12$\,dex, primarily due to over-recombination. This is 0.25\,dex lower than the value we adopt, although still within the uncertainties of the value from \citet{Grevesse:1998}, who reported an LTE value of $\log \epsilon_\odot(\mathrm{Rb}) = 2.60 \pm 0.10$.
In their Figure 3, \citet{korotin:20} presents the departure coefficients, $b = n_\mathrm{non-LTE}/n_\mathrm{LTE}$, for both the lower and upper levels of the near-IR transition at $\lambda15289$, which we use. In the solar atmosphere the lower level has a departure coefficient approximately half that of the ground state throughout the upper atmosphere, while the upper level shows minimal non-LTE effects. This, together with the fact that infrared lines generally are formed deeper in the atmosphere and are weaker than the resonance lines, suggests lower non-LTE corrections for the $\lambda15289$ line in the Sun. We can conclude that the non-LTE corrections behave similarly  in the Sun (having the same sign) and are likely smaller than those for the resonance lines. 

Furthermore, Figure 7 in the study by \citet{korotin:20} shows that the departure coefficients for the levels involved in the $\lambda15289$ transition are small in cool giants, such as those in our sample. A rough estimate, therefore, suggests that the non-LTE correction to the Rb abundance from $\lambda15289$ in giants should be small. However, detailed, star-by-star calculations are necessary to fully account for departures from LTE, including deviations in the source function.

Since we do not perform a non-LTE analysis, we also do not apply a non-LTE correction to the solar value to which we normalize. A future study of non-LTE effects in the near-IR lines is needed to resolve the $0.25$\,dex difference in the adopted solar reference values in different studies.

\section{Discussion}
\label{sec:discussion}

In this section, we discuss in which types of stars the Rb abundance can be measured from high-resolution near-IR spectra; how our Rb trend versus metallicity compares with other s-process elements and with trends reported in the literature based on optical spectra; and the current understanding of the cosmic production and Galactic chemical evolution of Rb.

\begin{figure*}
  \includegraphics[trim=0cm 0cm 0cm 0cm, clip, width=\linewidth]{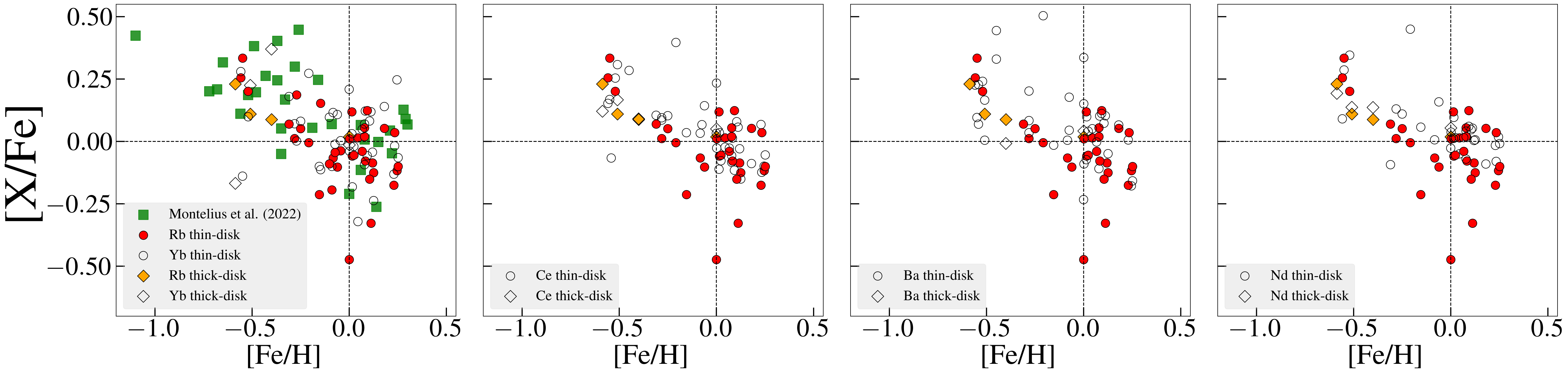}
  \caption{Determined [Rb/Fe] values (red circles and yellow diamonds) compared with the other s-process elements (open circles and diamonds) for the same set of stars from \citet{Nandakumar:24_21elements} for Yb, Ce, and Nd, and from \citet{Nandakumar:ba} for Ba. The [Yb/Fe] trend determined from IGRINS spectra of the warm K giants sample in \citet{montelius:22} is represented by the green filled squares. To facilitate comparison of trends, Yb and Ce trends have been shifted by -0.1 dex and +0.25 dex respectively to pass through the solar value. Thin- and thick-disk stars are marked with circles and diamonds, respectively.}
  \label{fig:Rb_sr_proc_trend}%
\end{figure*}

\begin{figure}
  \includegraphics[trim=0cm 0cm 0cm 0cm, clip, width=\linewidth]{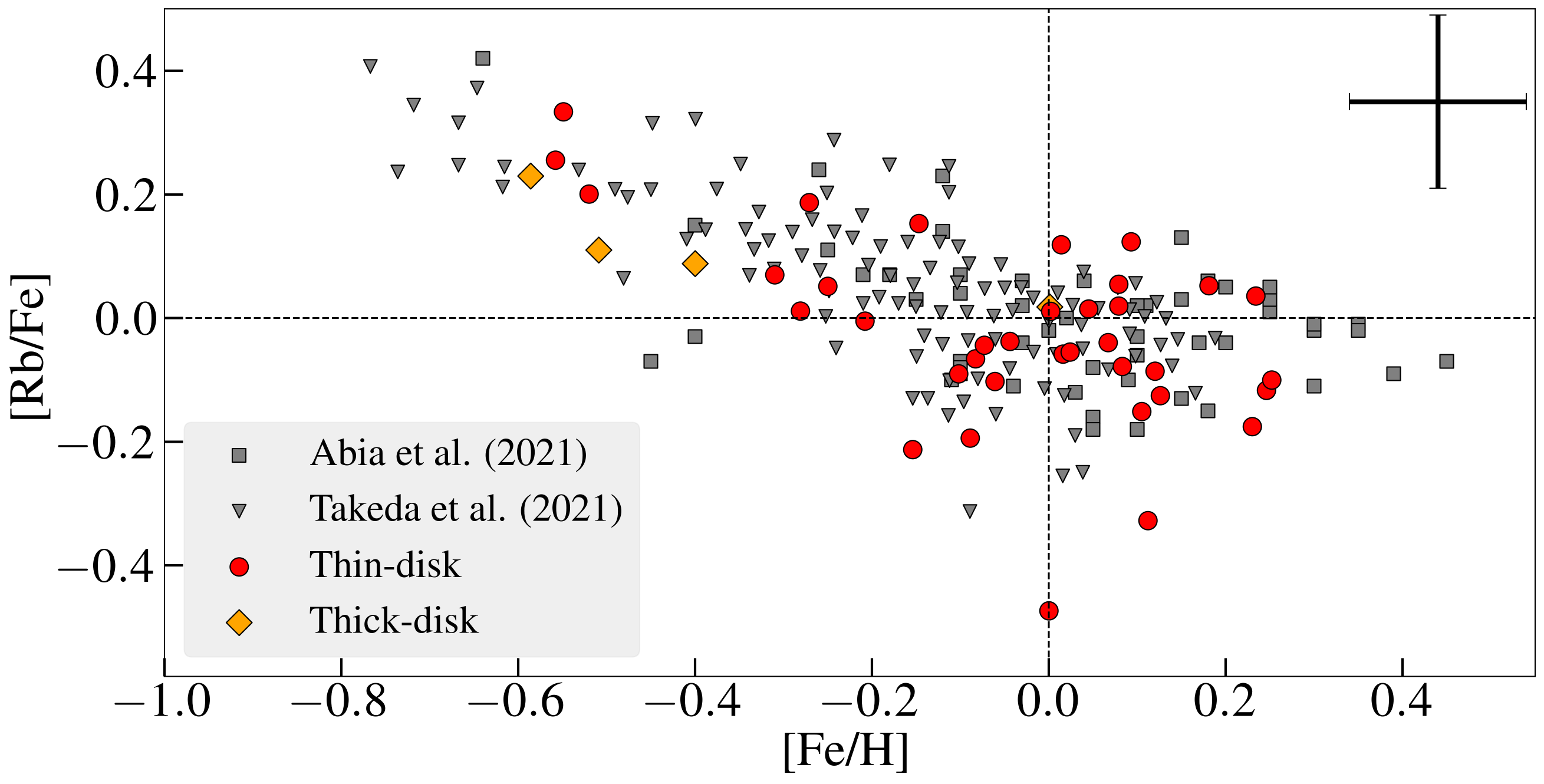}
  \caption{Our data set in red filled circles for the thin-disk stars and yellow diamonds for the thick-disk stars. The gray squares are from the study by \citet{abia:21} and the gray triangles are from the study by \citet{takeda:21}.}
  \label{fig:Rb_trend_literature}%
\end{figure}

\subsection{Detecting the  Rb\,{\sc i} $\lambda 15289$ Line in Stellar Spectra}

\begin{figure}
  \includegraphics[width=0.5\textwidth]{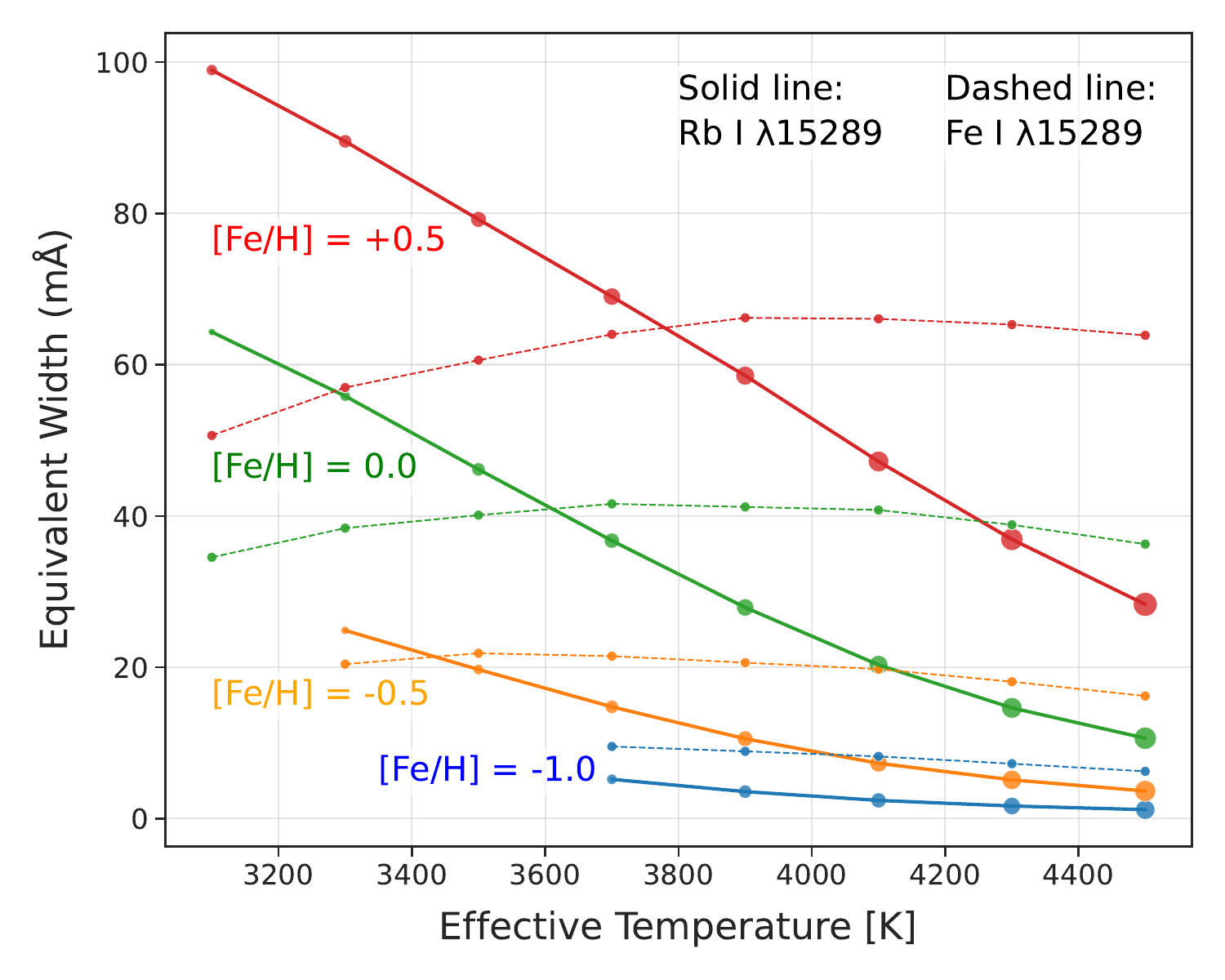}
  \caption{Equivalent widths in  m\AA\ of the Rubidium line (solid line) and the blending Fe line (dashed line) for typical red giants as a function of effective temperature of the stars. The colored lines show equal metallicities (blue for \feh\,$=-1.0$, orange for \feh\,$=-0.5$, green for \feh\,$=0.0$, and red for \feh\,$=+0.5$). The size of the symbols for the Rb abundances indicate the surface gravities corresponding to a given \teff\ and \feh\  according to a typical isochrone. The sizes go from largest being \logg\,$= 2.9$ to the smallest being \logg\,$= 0.0$. The equivalent width of the Rb\,\textsc{i}  line is quite insensitive to the surface gravity.}
  \label{fig:Rb_eqw}%
\end{figure}

With a first ionization potential as low as 4.2\,eV \citep{sansonetti:06}, most Rb is ionized, making Rb\,\textsc{i} a minority species throughout a stellar atmosphere. In Figure~\ref{fig:Rb_eqw}, we also see that the equivalent width of the Rb\,\textsc{i} line at $\lambda15289$ decreases rapidly with increasing temperature, where Rb is more strongly ionized. The line opacity of a minority species is sensitive to the electron pressure; however, this is also true for the continuous opacity in this wavelength region, which is dominated by H$^-_\mathrm{ff}$. In general, the equivalent widths of weak lines depends on the ratio of line to continuum opacity, $\chi_\mathrm{line}/\chi_\mathrm{cont}$. This implies that the sensitivity to electron pressure largely cancels out in this ratio, and hence in the line strength. This explains the very small changes in line strength with varying \logg, which affects the electron pressure through the hydrostatic equilibrium equation.

In Figure\,\ref{fig:Rb_eqw}, we also plot the equivalent width of the blending Fe\,\textsc{i} line, as well as that of the Rb\,\textsc{i} line, for typical red giant stars with realistic combinations of \teff ranging from 3100\,K to 3900\,K (in 200\,K intervals), \logg\ values from 0.0 to 1.7\,dex, and \feh values of $-1.0$, $-0.5$, $0.0$, and $+0.5$\,dex. From the Figure, we indeed see that the strength of the Rb\,\textsc{i} line decreases more rapidly with increasing temperature compared to the Fe\,\textsc{i} line, which remains relatively constant. This implies that the relative strength of the Rb\,\textsc{i} line is greatest at lower temperatures. For example, at \teff$<3400$\,K, the Rb\,{\sc i} line contributes more than half of the total spectral feature in giants with \feh$>-0.5$. Thus, not only does the Rb\,\textsc{i} line weaken with increasing temperature, but the Fe\,\textsc{i} blend also constitutes a larger fraction of the feature, making it increasingly difficult to derive a reliable Rb abundance for warmer stars. 

We also see in the Figure that the line strength naturally decreases with metallicity of the star. In our most metal-poor star, the thick-disk giant 2M18522108–3022143, with \teff$= 3578$\,K, \logg$= 0.45$, and \feh$=-0.6$, we still clearly detect the Rb line, with an equivalent width of approximately $18$\,m\AA.

Although the intrinsic line strengths of the blending Fe\,{\sc i} lines are well modeled, uncertainties in the Fe abundance may affect the residual line strength attributed to the Rb transition, due to incorrect strengths of the blending Fe lines. In cases where the Rb contribution is small relative to the Fe blend, this effect can be large. In Figure\,\ref{fig:Rb_trend}, the Rb abundances are color-coded by Rb line strength. We note that the lowest derived Rb abundances correspond to features with weak Rb lines, and are thus subject to larger uncertainties. For example, in the star 2M14322072-6215506 (with \teff=$3639$\,K and solar metallicity), which has the lowest measured [Rb/Fe] value of $-0.47$, the redward Fe blend is indeed overestimated in the synthesis, introducing additional uncertainty beyond that from the stellar parameters alone and leading to an underestimated Rb abundance. This underscores the large uncertainties associated with weak Rb line strengths. A further indication that the Rb abundance is underestimated comes from the corresponding Yb abundance in the same star. Since Rb and Yb share similar nucleosynthetic origin, both being produced in nearly equal parts by the s- and r-processes, the derived supersolar Yb abundance strongly suggests that the derived Rb abundance is too low.

\subsection{Comparison with other s-process Elements and Literature Trends}

Our [Rb/Fe] values as a function of metallicity for all stars in our sample are shown in Figure \ref{fig:Rb_trend}. In Figure\,\ref{fig:Rb_sr_proc_trend}, we compare this trend with those of other s-process elements (Yb, Ce, Ba, and Nd), derived for the same stars using the same spectra, as reported by \citet{Nandakumar:24_21elements,Nandakumar:ba}. The [Yb/Fe] trend determined from IGRINS spectra of the warm K giants in \citet{montelius:22} is also shown in the figure.  
To facilitate comparison of trends and their respective spreads, we normalize all trends to pass  through the solar value. Overall, we observe a general similarity among these elements, within the uncertainties. However, more data are needed to resolve finer differences between s-process elements with different fraction of the s- and r-process channels. Yb is the other element which is produced by both the r- and s-processes in roughly equal proportions in
the Solar system isotopic composition. No large differences can be seen in the [Rb/Fe] and the [Yb/Fe] trends, which is thus expected. 

In Figure\,\ref{fig:Rb_sr_proc_trend} there is also a star at \feh$=-0.21$ that is enhanced in s-process elements such as Ce, Ba, and Nd \citep[see also][]{Nandakumar:24_21elements,Nandakumar:ba}. However, it is not similarly enhanced in Yb or Rb, which may be expected if these elements are only partially produced in the s-process.
 
Early studies on the evolution of Rb as a function of metallicity in the Milky Way include \citet{gratton:94} and \citet{tomkin:99}. More recently, \citet{abia:21} presented Rb abundances for a sample of cool K- and M-type giants with metallicities in the range $-0.6<$\feh$0.4$, derived from resonance lines with modest non-LTE corrections. They found that [Rb/Fe] closely follows the trends of other s-process elements, such as [Zr/Fe] and [Sr/Fe]. We have included their [Rb/Fe] data in Figure\,\ref{fig:Rb_trend_literature} together with our [Rb/Fe] values. Similarly, \citet{takeda:21} derived non-LTE Rb abundances for giants within the same metallicity range, using the same resonance lines. These data are also shown in Figure\,\ref{fig:Rb_trend_literature}.
Our [Rb/Fe] trend with metallicity closely follows those found in these optical studies. This agreement demonstrates that we have successfully determined Rb abundances from the near-infrared line at $\lambda15289$, and it supports the assumption that non-LTE corrections for this line are indeed small.



\subsection{Cosmic Origin and Galactic Chemical Evolution of Rb}

\begin{figure}
  \includegraphics[trim=2cm 9.5cm 2cm 9cm, clip, width=\linewidth]
  {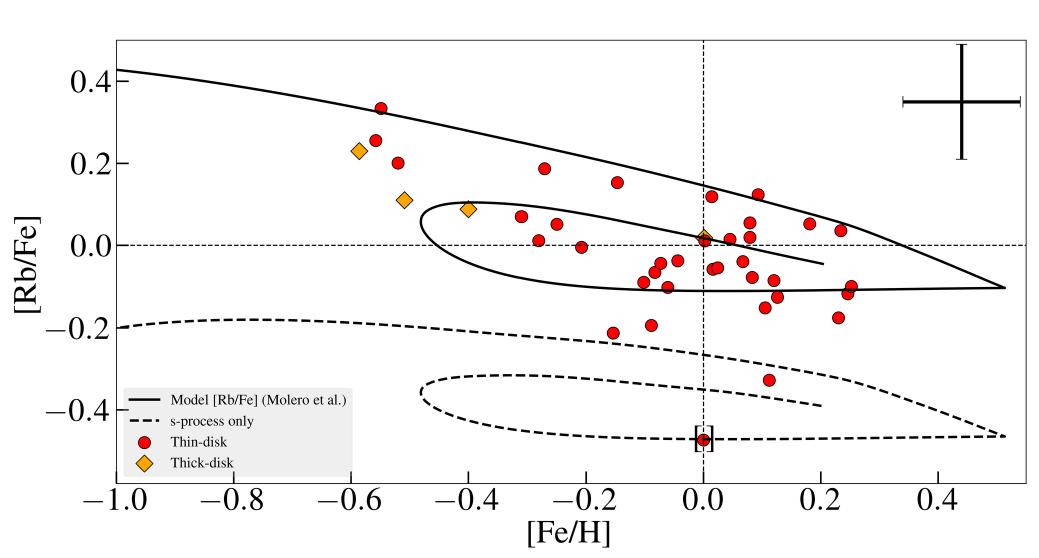}
  \caption{Our data plotted together with the results of our chemical evolution model under two scenarios: one in which Rb is synthesized exclusively through the s-process (dashed line), and another in which the contributions from both the s- and the r-process are included (full line). To ensure that these models pass through the solar value,  we have renormalized [Rb/Fe] by approximately $+0.2$\,dex. The lowest [Rb/Fe] abundance is highly uncertain, see text.  }
  \label{fig:Rb_GCE}%
\end{figure}

As a light first peak neutron-capture element, Rb has a complex nucleosynthetic origin involving contributions from both the r- and s-processes. As a consequence, its Galactic evolution reflects multiple production channels across different stellar sites.
From the r-process perspective, being close to the first peak, Rb may originate in part from the weak r-process that can occur in environments such as neutrino-driven winds from core-collapse supernovae, or possibly in magneto-rotational supernovae where rotation and magnetic fields enhance neutron capture. 

From the s-process side, \citet{prantzos:18} emphasized that both the weak s-process, active in rotating massive stars, and the main s-process, occurring in asymptotic giant branch (AGB) stars, play substantial roles in the production of Rb. The weak s-process, contributes to Rb early in the Galactic history. Meanwhile, the main s-process  becomes dominant at later times. In general, the production of Rb through the s-process depends sensitively on stellar mass, metallicity and, in the case of massive stars, rotation. 

To model the temporal evolution of Rb, we adopt the two-infall chemical evolution framework of \citet{Spitoni2019} for the solar neighborhood of the Milky Way, coupled with the nucleosynthesis prescriptions of \citet{Molero2025} (model 7). In this model, the r-process contribution to Rb production arises from both neutron star mergers (NSMs) and magneto-rotational supernovae (MR-SNe). NSMs are treated as a delayed r-process source, with their event rate dependent on the delay-time distribution (DTD) formalism of \citet{Simonetti2019}. The normalization of the NSM rate is constrained to match the most recent estimates from gravitational-wave observations (Abbott et al. 2021). The yields per merger event are scaled to match the Sr abundance inferred from the kilonova AT2017gfo, as discussed in Watson et al. (2019) and implemented in \citet{Molero2021}.
MR-SNe, in contrast, act as a prompt r-process source. They are assumed to originate from a small subset of massive stars (with progenitor masses in the range $10-20\ M_\odot$) and contribute r-process material according to the L0.75 model of \citet{Nishimura2017}.

As for the s-process, Rb has contributions from both a prompt and a delayed channel. The prompt s-process source is represented by rotating massive stars that explode as CC-SNe, with nucleosynthetic yields taken from the rotating models (set R) of \citet{Limongi2018}. Stellar rotation significantly enhances the s-process, particularly at low metallicities and for elements near the first s-process peak, such as Rb. Rather than assuming a uniform rotational velocity, we adopt the distribution of initial rotational velocities of \citet{prantzos:18}, which favors faster rotation at lower metallicity. This approach is supported by further studies (\citealp{Romano2019, Rizzuti2019, Molero2024}).
The delayed s-process component is attributed to asymptotic giant branch (AGB) stars, with yields taken from the extended sets of the FRUITY database (e.g., \citealp{Cristallo2015}).

In Figure\,\ref{fig:Rb_GCE}, we present the results of our chemical evolution model for [Rb/Fe], which includes contributions from both the s- and r-processes (full, black line). To ensure that the thin disk model passes through the solar value, as expected, we have increased the normalization by approximately 0.2\,dex. This underestimation may result from an insufficient production of Rb by massive stars, either from standard CC-SNe, MR-SNe, or a combination of both. Overall, the model agrees well with our data. In the range $-0.5<$\feh$<+0.2$, it predicts a spread in [Rb/Fe] due to the two-infall chemical evolution framework, which is consistent with our observations. The model also predicts that the thick disk should contain stars with higher [Rb/Fe] ratios than those in the thin disk at similar metallicities. Based on our data, and considering the abundance uncertainties shown in the Figure, we cannot rule out this general trend.

As noted above, Rb is produced in equal proportions by the s- and r-processes  in the solar isotope mixture (\citealp{prantzos:20}). The s-process component alone is insufficient to account for the full range of observational data, not only at solar, but also at low metallicities and in early Galactic environments. This limitation is clearly shown in Figure \ref{fig:Rb_GCE}, where we also present the results of our chemical evolution model in which Rb is synthesized exclusively through the s-process (the lower, dashed line). The model with only the s-process under-predicts the observed [Rb/Fe] ratios taking the uncertainties into account, reinforcing the need for at least one additional, prompt r-process source. Including r-process contributions from MR-SNe and NSMs brings the model predictions into much better agreement with the full observational dataset across the whole metallicity range. 

Our chemical evolution model for Yb, which, like Rb, is produced in roughly equal proportions by the r- and s-processes in the Solar system isotopic composition, shows a very similar trend to that of Rb, as expected. Ce, however, with an s/r ratio of 85/15, is predominantly produced by the s-process, and exhibits lower [Ce/Fe] values at low metallicities, followed by a rapid increase due to the delayed contribution from AGB stars. Although this contribution begins already at  \feh$\sim-1.0$, it becomes clearly visible in the overall trend only around 
\feh$\sim-0.5$. Our [Rb/Fe] data only barely extend to such low metallicities, but when combined with literature results \citep{abia:21,takeda:21}, they clearly demonstrate the need for  prompt-process contributions to Rb production, contributions not required to explain the behavior of Ce.

\section{Conclusions}
\label{sec:conclusion}

We have demonstrated that the  Rb\,\textsc{i} line at 15289.48\,\AA\ in the H-band is a reliable diagnostic of rubidium abundances in cool M giants, provided that high-resolution spectra ($R\sim45000$) with sufficient signal-to-noise are available. By carefully adjusting the line list, including blending Fe\,\textsc{i} lines as well as using the newest \loggf values for the Rb lines, we derived [Rb/Fe] ratios for 40 M giants in the solar neighborhood by performing detailed spectral synthesis. Our measurements reveal a decreasing trend of [Rb/Fe] with metallicity, consistent with the expected combined contributions of the s- and r-processes, and matching the behavior observed for Yb, another mixed-origin neutron-capture element.

Our results are in good agreement with previous optical studies based on the   Rb\,\textsc{i} resonance lines. The near-IR line at 15289.48\,\AA\  provides a robust alternative for abundance analysis, particularly in cool, dusty, or obscured populations where optical access is limited. Comparison with Galactic chemical evolution models further supports the need for both early and delayed sources of Rb, with contributions from rotating massive stars, AGB stars, neutron star mergers, and magneto-rotational supernovae.

With this study, Rb joins the growing list of neutron-capture elements measurable in high-resolution near-infrared spectra. This expands the chemical toolbox available for studies of the Milky Way's most reddened stellar populations, including those in the bulge and the Galactic Center with the Nuclear Stellar Disk and Nuclear Star Cluster.


\section*{Acknowledgments}
We thank the referee for their valuable comments and suggestions. N.R.\ acknowledge support from the Swedish Research Council (grant 2023-04744) and the Royal Physiographic Society in Lund through the Stiftelsen Walter Gyllenbergs and Märta och Erik Holmbergs donations.  G.N. acknowledges the support from the Crafoord Foundation via the Royal Swedish Academy of Sciences (Vetenskapsakademiens stiftelser; CR 2024-0034). H.H.\ acknowledges support from the Swedish Research Concil VR (grant 2023-05367). M.M. thanks the support from the Deutsche Forschungsgemeinschaft (DFG, German Research Foundation) – Project-ID 279384907 – SFB 1245, the State of Hessen within the Research Cluster ELEMENTS (Project ID 500/10.006).  H.J.\ acknowledges support from the Swedish Research Council VR (grant 2024-04989). 

This work used The Immersion Grating Infrared Spectrometer (IGRINS) was developed under a collaboration between the University of Texas at Austin and the Korea Astronomy and Space Science Institute (KASI) with the financial support of the US National Science Foundation under grants AST-1229522, AST-1702267 and AST-1908892, McDonald Observatory of the University of Texas at Austin, the Korean GMT Project of KASI, the Mt. Cuba Astronomical Foundation and Gemini Observatory.  The RRISA is maintained by the IGRINS Team with support from McDonald Observatory of the University of Texas at Austin and the US National Science Foundation under grant AST-1908892.

This work is based on observations obtained at the international Gemini Observatory, a program of NSF’s NOIRLab, which is managed by the Association of Universities for Research in Astronomy (AURA) under a cooperative agreement with the National Science Foundation on behalf of the Gemini Observatory partnership: the National Science Foundation (United States), National Research Council (Canada), Agencia Nacional de Investigaci\'{o}n y Desarrollo (Chile), Ministerio de Ciencia, Tecnolog\'{i}a e Innovaci\'{o}n (Argentina), Minist\'{e}rio da Ci\^{e}ncia, Tecnologia, Inova\c{c}\~{o}es e Comunica\c{c}\~{o}es (Brazil), and Korea Astronomy and Space Science Institute (Republic of Korea).
The following software and programming languages made this
research possible: TOPCAT (version 4.6; \citealt{topcat}); Python (version 3.8) and its packages ASTROPY (version 5.0; \citealt{astropy}), SCIPY \citep{scipy}, MATPLOTLIB \citep{matplotlib} and NUMPY \citep{numpy}.

%
%



\begin{thebibliography}{}
\expandafter\ifx\csname natexlab\endcsname\relax\def\natexlab#1{#1}\fi
\providecommand{\url}[1]{\href{#1}{#1}}
\providecommand{\dodoi}[1]{doi:~\href{http://doi.org/#1}{\nolinkurl{#1}}}
\providecommand{\doeprint}[1]{\href{http://ascl.net/#1}{\nolinkurl{http://ascl.net/#1}}}
\providecommand{\doarXiv}[1]{\href{https://arxiv.org/abs/#1}{\nolinkurl{https://arxiv.org/abs/#1}}}

\bibitem[{{Abdurro'uf} {et~al.}(2022){Abdurro'uf}, {Accetta}, {Aerts}, {Silva Aguirre}, {Ahumada}, {Ajgaonkar}, {Filiz Ak}, {Alam}, {Allende Prieto}, {Almeida}, {Anders}, {Anderson}, {Andrews}, {Anguiano}, {Aquino-Ort{\'\i}z}, {Arag{\'o}n-Salamanca}, {Argudo-Fern{\'a}ndez}, {Ata}, {Aubert}, {Avila-Reese}, {Badenes}, {Barb{\'a}}, {Barger}, {Barrera-Ballesteros}, {Beaton}, {Beers}, {Belfiore}, {Bender}, {Bernardi}, {Bershady}, {Beutler}, {Bidin}, {Bird}, {Bizyaev}, {Blanc}, {Blanton}, {Boardman}, {Bolton}, {Boquien}, {Borissova}, {Bovy}, {Brandt}, {Brown}, {Brownstein}, {Brusa}, {Buchner}, {Bundy}, {Burchett}, {Bureau}, {Burgasser}, {Cabang}, {Campbell}, {Cappellari}, {Carlberg}, {Wanderley}, {Carrera}, {Cash}, {Chen}, {Chen}, {Cherinka}, {Chiappini}, {Choi}, {Chojnowski}, {Chung}, {Clerc}, {Cohen}, {Comerford}, {Comparat}, {da Costa}, {Covey}, {Crane}, {Cruz-Gonzalez}, {Culhane}, {Cunha}, {Dai}, {Damke}, {Darling}, {Davidson}, {Davies}, {Dawson}, {De Lee}, {Diamond-Stanic}, {Cano-D{\'\i}az}, {S{\'a}nchez},
  {Donor}, {Duckworth}, {Dwelly}, {Eisenstein}, {Elsworth}, {Emsellem}, {Eracleous}, {Escoffier}, {Fan}, {Farr}, {Feng}, {Fern{\'a}ndez-Trincado}, {Feuillet}, {Filipp}, {Fillingham}, {Frinchaboy}, {Fromenteau}, {Galbany}, {Garc{\'\i}a}, {Garc{\'\i}a-Hern{\'a}ndez}, {Ge}, {Geisler}, {Gelfand}, {G{\'e}ron}, {Gibson}, {Goddy}, {Godoy-Rivera}, {Grabowski}, {Green}, {Greener}, {Grier}, {Griffith}, {Guo}, {Guy}, {Hadjara}, {Harding}, {Hasselquist}, {Hayes}, {Hearty}, {Hern{\'a}ndez}, {Hill}, {Hogg}, {Holtzman}, {Horta}, {Hsieh}, {Hsu}, {Hsu}, {Huber}, {Huertas-Company}, {Hutchinson}, {Hwang}, {Ibarra-Medel}, {Chitham}, {Ilha}, {Imig}, {Jaekle}, {Jayasinghe}, {Ji}, {Johnson}, {Jones}, {J{\"o}nsson}, {Katkov}, {Khalatyan}, {Kinemuchi}, {Kisku}, {Knapen}, {Kneib}, {Kollmeier}, {Kong}, {Kounkel}, {Kreckel}, {Krishnarao}, {Lacerna}, {Lane}, {Langgin}, {Lavender}, {Law}, {Lazarz}, {Leung}, {Leung}, {Lewis}, {Li}, {Li}, {Lian}, {Liang}, {Lin}, {Lin}, {Lin}, {Lintott}, {Long}, {Longa-Pe{\~n}a}, {L{\'o}pez-Cob{\'a}}, {Lu},
  {Lundgren}, {Luo}, {Mackereth}, {de la Macorra}, {Mahadevan}, {Majewski}, {Manchado}, {Mandeville}, {Maraston}, {Margalef-Bentabol}, {Masseron}, {Masters}, {Mathur}, {McDermid}, {Mckay}, {Merloni}, {Merrifield}, {Meszaros}, {Miglio}, {Di Mille}, {Minniti}, {Minsley}, \& {Monachesi}}]{apogee_dr17}
{Abdurro'uf}, {Accetta}, K., {Aerts}, C., {et~al.} 2022, \apjs, 259, 35, \dodoi{10.3847/1538-4365/ac4414}

\bibitem[{{Abia} {et~al.}(2021){Abia}, {de Laverny}, {Korotin}, {Asensio Ramos}, {Recio-Blanco}, \& {Prantzos}}]{abia:21}
{Abia}, C., {de Laverny}, P., {Korotin}, S., {et~al.} 2021, \aap, 648, A107, \dodoi{10.1051/0004-6361/202040250}

\bibitem[{{Af{\c{s}}ar} {et~al.}(2016){Af{\c{s}}ar}, {Sneden}, {Frebel}, {Kim}, {Mace}, {Kaplan}, {Lee}, {Oh}, {Sok Oh}, {Pak}, {Park}, {Pavel}, {Yuk}, \& {Jaffe}}]{afsar:16}
{Af{\c{s}}ar}, M., {Sneden}, C., {Frebel}, A., {et~al.} 2016, \apj, 819, 103, \dodoi{10.3847/0004-637X/819/2/103}

\bibitem[{{Af{\c{s}}ar} {et~al.}(2018){Af{\c{s}}ar}, {Sneden}, {Wood}, {Lawler}, {Bozkurt}, {B{\"o}cek Topcu}, {Mace}, {Kim}, \& {Jaffe}}]{Afsar:2018}
{Af{\c{s}}ar}, M., {Sneden}, C., {Wood}, M.~P., {et~al.} 2018, \apj, 865, 44, \dodoi{10.3847/1538-4357/aada0c}

\bibitem[{{Amarsi} {et~al.}(2019){Amarsi}, {Nissen}, \& {Sk{\'u}lad{\'o}ttir}}]{Amarsi:2019}
{Amarsi}, A.~M., {Nissen}, P.~E., \& {Sk{\'u}lad{\'o}ttir}, {\'A}. 2019, \aap, 630, A104, \dodoi{10.1051/0004-6361/201936265}

\bibitem[{{Asplund} {et~al.}(2009){Asplund}, {Grevesse}, {Sauval}, \& {Scott}}]{Asplund:2009}
{Asplund}, M., {Grevesse}, N., {Sauval}, A.~J., \& {Scott}, P. 2009, \araa, 47, 481, \dodoi{10.1146/annurev.astro.46.060407.145222}

\bibitem[{{Astropy Collaboration} {et~al.}(2022){Astropy Collaboration}, {Price-Whelan}, {Lim}, {Earl}, {Starkman}, {Bradley}, {Shupe}, {Patil}, {Corrales}, {Brasseur}, {N{\"o}the}, {Donath}, {Tollerud}, {Morris}, {Ginsburg}, {Vaher}, {Weaver}, {Tocknell}, {Jamieson}, {van Kerkwijk}, {Robitaille}, {Merry}, {Bachetti}, {G{\"u}nther}, {Aldcroft}, {Alvarado-Montes}, {Archibald}, {B{\'o}di}, {Bapat}, {Barentsen}, {Baz{\'a}n}, {Biswas}, {Boquien}, {Burke}, {Cara}, {Cara}, {Conroy}, {Conseil}, {Craig}, {Cross}, {Cruz}, {D'Eugenio}, {Dencheva}, {Devillepoix}, {Dietrich}, {Eigenbrot}, {Erben}, {Ferreira}, {Foreman-Mackey}, {Fox}, {Freij}, {Garg}, {Geda}, {Glattly}, {Gondhalekar}, {Gordon}, {Grant}, {Greenfield}, {Groener}, {Guest}, {Gurovich}, {Handberg}, {Hart}, {Hatfield-Dodds}, {Homeier}, {Hosseinzadeh}, {Jenness}, {Jones}, {Joseph}, {Kalmbach}, {Karamehmetoglu}, {Ka{\l}uszy{\'n}ski}, {Kelley}, {Kern}, {Kerzendorf}, {Koch}, {Kulumani}, {Lee}, {Ly}, {Ma}, {MacBride}, {Maljaars}, {Muna}, {Murphy}, {Norman},
  {O'Steen}, {Oman}, {Pacifici}, {Pascual}, {Pascual-Granado}, {Patil}, {Perren}, {Pickering}, {Rastogi}, {Roulston}, {Ryan}, {Rykoff}, {Sabater}, {Sakurikar}, {Salgado}, {Sanghi}, {Saunders}, {Savchenko}, {Schwardt}, {Seifert-Eckert}, {Shih}, {Jain}, {Shukla}, {Sick}, {Simpson}, {Singanamalla}, {Singer}, {Singhal}, {Sinha}, {Sip{\H{o}}cz}, {Spitler}, {Stansby}, {Streicher}, {{\v{S}}umak}, {Swinbank}, {Taranu}, {Tewary}, {Tremblay}, {Val-Borro}, {Van Kooten}, {Vasovi{\'c}}, {Verma}, {de Miranda Cardoso}, {Williams}, {Wilson}, {Winkel}, {Wood-Vasey}, {Xue}, {Yoachim}, {Zhang}, {Zonca}, \& {Astropy Project Contributors}}]{astropy}
{Astropy Collaboration}, {Price-Whelan}, A.~M., {Lim}, P.~L., {et~al.} 2022, \apj, 935, 167, \dodoi{10.3847/1538-4357/ac7c74}

\bibitem[{{Bisterzo} {et~al.}(2014){Bisterzo}, {Travaglio}, {Gallino}, {Wiescher}, \& {K{\"a}ppeler}}]{bisterzo:14}
{Bisterzo}, S., {Travaglio}, C., {Gallino}, R., {Wiescher}, M., \& {K{\"a}ppeler}, F. 2014, \apj, 787, 10, \dodoi{10.1088/0004-637X/787/1/10}

\bibitem[{{B{\"o}cek Topcu} {et~al.}(2019){B{\"o}cek Topcu}, {Af{\c{s}}ar}, {Sneden}, {Pilachowski}, {Denissenkov}, {VandenBerg}, {Strickland}, {{\"O}zdemir}, {Mace}, {Kim}, \& {Jaffe}}]{bocek:19}
{B{\"o}cek Topcu}, G., {Af{\c{s}}ar}, M., {Sneden}, C., {et~al.} 2019, \mnras, 485, 4625, \dodoi{10.1093/mnras/stz727}

\bibitem[{{B{\"o}cek Topcu} {et~al.}(2020){B{\"o}cek Topcu}, {Af{\c{s}}ar}, {Sneden}, {Pilachowski}, {Denissenkov}, {VandenBerg}, {Wright}, {Mace}, {Jaffe}, {Strickland}, {Kim}, \& {Sokal}}]{bocek:20}
---. 2020, \mnras, 491, 544, \dodoi{10.1093/mnras/stz3008}

\bibitem[{{Brady} {et~al.}(2023){Brady}, {Sneden}, {Pilachowski}, {Af{\c{s}}ar}, {Mace}, {Jaffe}, {Gosnell}, \& {Seifert}}]{brady:23}
{Brady}, K.~E., {Sneden}, C., {Pilachowski}, C.~A., {et~al.} 2023, \aj, 166, 154, \dodoi{10.3847/1538-3881/acf2f3}

\bibitem[{{Cox} {et~al.}(2014){Cox}, {Cami}, {Kaper}, {Ehrenfreund}, {Foing}, {Ochsendorf}, {van Hooff}, \& {Salama}}]{dib:cox}
{Cox}, N.~L.~J., {Cami}, J., {Kaper}, L., {et~al.} 2014, \aap, 569, A117, \dodoi{10.1051/0004-6361/201323061}

\bibitem[{{Cristallo} {et~al.}(2015){Cristallo}, {Straniero}, {Piersanti}, \& {Gobrecht}}]{Cristallo2015}
{Cristallo}, S., {Straniero}, O., {Piersanti}, L., \& {Gobrecht}, D. 2015, \apjs, 219, 40, \dodoi{10.1088/0067-0049/219/2/40}

\bibitem[{{Demarque} {et~al.}(2004){Demarque}, {Woo}, {Kim}, \& {Yi}}]{Demarque:2004}
{Demarque}, P., {Woo}, J.-H., {Kim}, Y.-C., \& {Yi}, S.~K. 2004, \apjs, 155, 667, \dodoi{10.1086/424966}

\bibitem[{{Ebenbichler} {et~al.}(2022){Ebenbichler}, {Postel}, {Przybilla}, {Seifahrt}, {We{\ss}mayer}, {Kausch}, {Firnstein}, {Butler}, {Kaufer}, \& {Linnartz}}]{dib_crires}
{Ebenbichler}, A., {Postel}, A., {Przybilla}, N., {et~al.} 2022, \aap, 662, A81, \dodoi{10.1051/0004-6361/202142990}

\bibitem[{{Elyajouri} {et~al.}(2017){Elyajouri}, {Lallement}, {Monreal-Ibero}, {Capitanio}, \& {Cox}}]{dib:elya}
{Elyajouri}, M., {Lallement}, R., {Monreal-Ibero}, A., {Capitanio}, L., \& {Cox}, N.~L.~J. 2017, \aap, 600, A129, \dodoi{10.1051/0004-6361/201630088}

\bibitem[{{Feinberg} {et~al.}(2024){Feinberg}, {Ziemer}, {Ansdell}, {Crooke}, {Dressing}, {Mennesson}, {O'Meara}, {Pepper}, \& {Roberge}}]{HWO}
{Feinberg}, L., {Ziemer}, J., {Ansdell}, M., {et~al.} 2024, in Society of Photo-Optical Instrumentation Engineers (SPIE) Conference Series, Vol. 13092, Space Telescopes and Instrumentation 2024: Optical, Infrared, and Millimeter Wave, ed. L.~E. {Coyle}, S.~{Matsuura}, \& M.~D. {Perrin}, 130921N, \dodoi{10.1117/12.3018328}

\bibitem[{{Galazutdinov} {et~al.}(2017){Galazutdinov}, {Lee}, {Han}, {Lee}, {Valyavin}, \& {Kre{\l}owski}}]{dib:galazut}
{Galazutdinov}, G.~A., {Lee}, J.-J., {Han}, I., {et~al.} 2017, \mnras, 467, 3099, \dodoi{10.1093/mnras/stx330}

\bibitem[{{Garc{\'\i}a-Hern{\'a}ndez} {et~al.}(2023){Garc{\'\i}a-Hern{\'a}ndez}, {Rao}, {Lambert}, {Eriksson}, {Reddy}, \& {Masseron}}]{garcia:23}
{Garc{\'\i}a-Hern{\'a}ndez}, D.~A., {Rao}, N.~K., {Lambert}, D.~L., {et~al.} 2023, \apj, 948, 15, \dodoi{10.3847/1538-4357/acc574}

\bibitem[{{Geballe}(2016)}]{dib:geballe1}
{Geballe}, T.~R. 2016, in Journal of Physics Conference Series, Vol. 728, Journal of Physics Conference Series, 062005, \dodoi{10.1088/1742-6596/728/6/062005}

\bibitem[{{Geballe} {et~al.}(2011){Geballe}, {Najarro}, {Figer}, {Schlegelmilch}, \& {de La Fuente}}]{dib:geballe_nature}
{Geballe}, T.~R., {Najarro}, F., {Figer}, D.~F., {Schlegelmilch}, B.~W., \& {de La Fuente}, D. 2011, \nat, 479, 200, \dodoi{10.1038/nature10527}

\bibitem[{{Gonzalez} {et~al.}(2020){Gonzalez}, {Mucciarelli}, {Origlia}, {Schultheis}, {Caffau}, {Di Matteo}, {Randich}, {Recio-Blanco}, {Zoccali}, {Bonifacio}, {Dalessandro}, {Schiavon}, {Pancino}, {Taylor}, {Valenti}, {Rojas-Arriagada}, {Sacco}, {Biazzo}, {Bellazzini}, {Cioni}, {Clementini}, {Contreras Ramos}, {de Laverny}, {Evans}, {Haywood}, {Hill}, {Ibata}, {Lucatello}, {Magrini}, {Martin}, {Nisini}, {Sanna}, {Cirasuolo}, {Maiolino}, {Afonso}, {Lilly}, {Flores}, {Oliva}, {Paltani}, \& {Vanzi}}]{MOONS2020}
{Gonzalez}, O.~A., {Mucciarelli}, A., {Origlia}, L., {et~al.} 2020, The Messenger, 180, 18, \dodoi{10.18727/0722-6691/5196}

\bibitem[{{Gratton} \& {Sneden}(1994)}]{gratton:94}
{Gratton}, R.~G., \& {Sneden}, C. 1994, \aap, 287, 927

\bibitem[{{Grevesse} \& {Sauval}(1998)}]{Grevesse:1998}
{Grevesse}, N., \& {Sauval}, A.~J. 1998, \ssr, 85, 161, \dodoi{10.1023/A:1005161325181}

\bibitem[{{Grisoni} {et~al.}(2020){Grisoni}, {Romano}, {Spitoni}, {Matteucci}, {Ryde}, \& {J{\"o}nsson}}]{Grisoni:2020}
{Grisoni}, V., {Romano}, D., {Spitoni}, E., {et~al.} 2020, \mnras, 498, 1252, \dodoi{10.1093/mnras/staa2316}

\bibitem[{{Gully-Santiago} {et~al.}(2012){Gully-Santiago}, {Wang}, {Deen}, \& {Jaffe}}]{Gully:2012}
{Gully-Santiago}, M., {Wang}, W., {Deen}, C., \& {Jaffe}, D. 2012, in Society of Photo-Optical Instrumentation Engineers (SPIE) Conference Series, Vol. 8450, Modern Technologies in Space- and Ground-based Telescopes and Instrumentation II, ed. R.~{Navarro}, C.~R. {Cunningham}, \& E.~{Prieto}, 84502S, \dodoi{10.1117/12.926434}

\bibitem[{{Gustafsson} {et~al.}(2008){Gustafsson}, {Edvardsson}, {Eriksson}, {et~al.}}]{marcs:08}
{Gustafsson}, B., {Edvardsson}, B., {Eriksson}, K., {et~al.} 2008, \aap, 486, 951

\bibitem[{{Hayes} {et~al.}(2022){Hayes}, {Masseron}, {Sobeck}, {Garcia-Hernandez}, {Allende Prieto}, {Beaton}, {Cunha}, {Hasselquist}, {Holtzman}, {Jonsson}, {Majewski}, {Shetrone}, {Smith}, \& {Almeida}}]{Hayes:2022}
{Hayes}, C.~R., {Masseron}, T., {Sobeck}, J., {et~al.} 2022, arXiv e-prints, arXiv:2208.00071.
\newblock \doarXiv{2208.00071}

\bibitem[{{Holanda} {et~al.}(2024){Holanda}, {Roriz}, {Drake}, {Junqueira}, {Daflon}, {da Silva}, \& {Pereira}}]{holanda:24}
{Holanda}, N., {Roriz}, M.~P., {Drake}, N.~A., {et~al.} 2024, \mnras, 527, 1389, \dodoi{10.1093/mnras/stad3280}

\bibitem[{{Hunter}(2007)}]{matplotlib}
{Hunter}, J.~D. 2007, Computing in Science and Engineering, 9, 90, \dodoi{10.1109/MCSE.2007.55}

\bibitem[{{Jeong} {et~al.}(2014){Jeong}, {Chun}, {Oh}, {Park}, {Yuk}, {Oh}, {Kim}, {Ko}, {Pavel}, {Yu}, \& {Jaffe}}]{Jeong:2014}
{Jeong}, U., {Chun}, M.-Y., {Oh}, J.~S., {et~al.} 2014, in Society of Photo-Optical Instrumentation Engineers (SPIE) Conference Series, Vol. 9154, High Energy, Optical, and Infrared Detectors for Astronomy VI, ed. A.~D. {Holland} \& J.~{Beletic}, 91541X, \dodoi{10.1117/12.2055589}

\bibitem[{Johansson(1961)}]{Johansson:61}
Johansson, I. 1961, Ark. Fys., 20, 135

\bibitem[{Kaplan {et~al.}(2024)Kaplan, Lee, Sawczynec, \& Kim}]{kaplan_plp}
Kaplan, K., Lee, J.-J., Sawczynec, E., \& Kim, H.-J. 2024, igrins/plp, 3.0.0,  Zenodo, \dodoi{10.5281/zenodo.11080095}

\bibitem[{{Korotin}(2020)}]{korotin:20}
{Korotin}, S.~A. 2020, Astronomy Letters, 46, 541, \dodoi{10.1134/S1063773720080022}

\bibitem[{Kramida {et~al.}(2024)Kramida, {Yu.~Ralchenko}, Reader, \& {and NIST ASD Team}}]{NIST_ASD}
Kramida, A., {Yu.~Ralchenko}, Reader, J., \& {and NIST ASD Team}. 2024, {NIST Atomic Spectra Database (ver. 5.12), [Online]. Available: {\tt{https://physics.nist.gov/asd}} [2025, May 2]. National Institute of Standards and Technology, Gaithersburg, MD.}

\bibitem[{{Kurucz}(2014)}]{K14}
{Kurucz}, R.~L. 2014, Robert L. Kurucz on-line database of observed and predicted atomic transitions

\bibitem[{{Limongi} \& {Chieffi}(2018)}]{Limongi2018}
{Limongi}, M., \& {Chieffi}, A. 2018, \apjs, 237, 13, \dodoi{10.3847/1538-4365/aacb24}

\bibitem[{{Mace} {et~al.}(2016){Mace}, {Kim}, {Jaffe}, {Park}, {Lee}, {Kaplan}, {Yu}, {Yuk}, {Chun}, {Pak}, {Kim}, {Lee}, {Sneden}, {Afsar}, {Pavel}, {Lee}, {Oh}, {Jeong}, {Park}, {Kidder}, {Lee}, {Nguyen Le}, {McLane}, {Gully-Santiago}, {Oh}, {Lee}, {Hwang}, \& {Park}}]{Mace:2016}
{Mace}, G., {Kim}, H., {Jaffe}, D.~T., {et~al.} 2016, in Society of Photo-Optical Instrumentation Engineers (SPIE) Conference Series, Vol. 9908, Ground-based and Airborne Instrumentation for Astronomy VI, ed. C.~J. {Evans}, L.~{Simard}, \& H.~{Takami}, 99080C, \dodoi{10.1117/12.2232780}

\bibitem[{{Mace} {et~al.}(2018){Mace}, {Sokal}, {Lee}, {Oh}, {Park}, {Lee}, {Good}, {MacQueen}, {Oh}, {Kaplan}, {Kidder}, {Chun}, {Yuk}, {Jeong}, {Pak}, {Kim}, {Nah}, {Lee}, {Yu}, {Hwang}, {Park}, {Kim}, {Chinn}, {Peck}, {Diaz}, {Rutten}, {Prato}, {Jacoby}, {Cornelius}, {Hardesty}, {DeGroff}, {Dunham}, {Levine}, {Nofi}, {Lopez-Valdivia}, {Weinberger}, \& {Jaffe}}]{Mace:2018}
{Mace}, G., {Sokal}, K., {Lee}, J.-J., {et~al.} 2018, in Society of Photo-Optical Instrumentation Engineers (SPIE) Conference Series, Vol. 10702, Ground-based and Airborne Instrumentation for Astronomy VII, ed. C.~J. {Evans}, L.~{Simard}, \& H.~{Takami}, 107020Q, \dodoi{10.1117/12.2312345}

\bibitem[{{Majewski} {et~al.}(2017){Majewski}, {Schiavon}, {Frinchaboy}, {Allende Prieto}, {Barkhouser}, {Bizyaev}, {Blank}, {Brunner}, {Burton}, {Carrera}, {Chojnowski}, {Cunha}, {Epstein}, {Fitzgerald}, {Garc{\'\i}a P{\'e}rez}, {Hearty}, {Henderson}, {Holtzman}, {Johnson}, {Lam}, {Lawler}, {Maseman}, {M{\'e}sz{\'a}ros}, {Nelson}, {Nguyen}, {Nidever}, {Pinsonneault}, {Shetrone}, {Smee}, {Smith}, {Stolberg}, {Skrutskie}, {Walker}, {Wilson}, {Zasowski}, {Anders}, {Basu}, {Beland}, {Blanton}, {Bovy}, {Brownstein}, {Carlberg}, {Chaplin}, {Chiappini}, {Eisenstein}, {Elsworth}, {Feuillet}, {Fleming}, {Galbraith-Frew}, {Garc{\'\i}a}, {Garc{\'\i}a-Hern{\'a}ndez}, {Gillespie}, {Girardi}, {Gunn}, {Hasselquist}, {Hayden}, {Hekker}, {Ivans}, {Kinemuchi}, {Klaene}, {Mahadevan}, {Mathur}, {Mosser}, {Muna}, {Munn}, {Nichol}, {O'Connell}, {Parejko}, {Robin}, {Rocha-Pinto}, {Schultheis}, {Serenelli}, {Shane}, {Silva Aguirre}, {Sobeck}, {Thompson}, {Troup}, {Weinberg}, \& {Zamora}}]{apogee}
{Majewski}, S.~R., {Schiavon}, R.~P., {Frinchaboy}, P.~M., {et~al.} 2017, \aj, 154, 94, \dodoi{10.3847/1538-3881/aa784d}

\bibitem[{{Manea} {et~al.}(2024){Manea}, {Hawkins}, {Ness}, {Buder}, {Martell}, \& {Zucker}}]{manea:23}
{Manea}, C., {Hawkins}, K., {Ness}, M.~K., {et~al.} 2024, \apj, 972, 69, \dodoi{10.3847/1538-4357/ad58d9}

\bibitem[{{Marconi} {et~al.}(2024){Marconi}, {Abreu}, {Adibekyan}, {Alberti}, {Albrecht}, {Alcaniz}, {Aliverti}, {Allende Prieto}, {Alvarado-Gomez}, {Alves}, {Amado}, {Amate}, {Andersen}, {Antoniucci}, {Artigau}, {Bailet}, {Baker}, {Baldini}, {Balestra}, {Barnes}, {Baron}, {Barros}, {Bauer}, {Beaulieu}, {Bellido-Tirado}, {Benneke}, {Bensby}, {Bergin}, {Berio}, {Biazzo}, {Bigot}, {Bik}, {Birkby}, {Blind}, {Boebion}, {Boisse}, {Bolmont}, {Bolton}, {Bonaglia}, {Bonfils}, {Bonhomme}, {Borsa}, {Bouret}, {Brandeker}, {Brandner}, {Broeg}, {Brogi}, {Brousseau}, {Brucalassi}, {Brynnel}, {Buchhave}, {Buscher}, {Cabona}, {Cabral}, {Calderone}, {Calvo-Ortega}, {Cantalloube}, {Canto Martins}, {Carbonaro}, {Caujolle}, {Chauvin}, {Chazelas}, {Cheffot}, {Cheng}, {Chiavassa}, {Christensen}, {Cirami}, {Cirasuolo}, {Cook}, {Cooke}, {Coretti}, {Covino}, {Cowan}, {Cresci}, {Cristiani}, {Cunha Parro}, {Cupani}, {D'Odorico}, {Dadi}, {de Castro Le{\~a}o}, {De Cia}, {De Medeiros}, {Debras}, {Debus}, {Delorme}, {Demangeon}, {Derie},
  {Dessauges-Zavadsky}, {Di Marcantonio}, {Di Stefano}, {Dionies}, {Domiciano de Souza}, {Doyon}, {Dunn}, {Egner}, {Ehrenreich}, {Faria}, {Ferruzzi}, {Feruglio}, {Fisher}, {Fontana}, {Frank}, {Fuesslein}, {Fumagalli}, {Fusco}, {Fynbo}, {Gabella}, {Gaessler}, {Gallo}, {Gao}, {Genolet}, {Genoni}, {Giacobbe}, {Giro}, {Gon{\c{c}}alves}, {Gonzalez}, {Gonz{\'a}lez-Hern{\'a}ndez}, {Gouvret}, {Gracia T{\'e}mich}, {Haehnelt}, {Haniff}, {Hatzes}, {Helled}, {Hoeijmakers}, {Hughes}, {Huke}, {Ivanisenko}, {J{\"a}rvinen}, {J{\"a}rvinen}, {Kaminski}, {Kern}, {Knoche}, {Kordt}, {Korhonen}, {Korn}, {Kouach}, {Kowzan}, {Kreidberg}, {Landoni}, {Lanotte}, {Lavail}, {Lavie}, {Lee}, {Lehmitz}, {Li}, {Li}, {Liske}, {Lovis}, {Lucatello}, {Lunney}, {MacIntosh}, {Madhusudhan}, {Magrini}, {Maiolino}, {Maldonado}, {Malo}, {Man}, {Marquart}, {Marques}, {Marques}, {Martinez}, {Martins}, {Martins}, {Martins}, {Maslowski}, {Mason}, {Mason}, {McCracken}, {Melo e Sousa}, {Mergo}, {Micela}, {Milakovi{\'c}}, {Molli{\`e}re}, {Monteiro},
  {Montgomery}, {Mordasini}, {Morin}, {Mucciarelli}, {Murphy}, {N'Diaye}, {Nardetto}, {Neichel}, {Neri}, {Niedzielski}, {Niemczura}, {Nisini}, {Nortmann}, {Noterdaeme}, {Nunes}, {Oggioni}, {Olchewsky}, {Oliva}, {{\"O}nel}, {Origlia}, {{\"O}stlin}, {Ouellette}, {Pall{\'e}}, {Papaderos}, {Pariani}, \& {Pasquini}}]{Andes}
{Marconi}, A., {Abreu}, M., {Adibekyan}, V., {et~al.} 2024, in Society of Photo-Optical Instrumentation Engineers (SPIE) Conference Series, Vol. 13096, Ground-based and Airborne Instrumentation for Astronomy X, ed. J.~J. {Bryant}, K.~{Motohara}, \& J.~R.~D. {Vernet}, 1309613, \dodoi{10.1117/12.3017966}

\bibitem[{{Matteucci}(2012)}]{matteucci:12}
{Matteucci}, F. 2012, {Chemical Evolution of Galaxies}, \dodoi{10.1007/978-3-642-22491-1}

\bibitem[{{Matteucci}(2021)}]{Matteucci:2021}
---. 2021, \aapr, 29, 5, \dodoi{10.1007/s00159-021-00133-8}

\bibitem[{{Mead} {et~al.}(2025){Mead}, {De La Garza}, \& {Ness}}]{mead}
{Mead}, J., {De La Garza}, R., \& {Ness}, M. 2025, arXiv e-prints, arXiv:2504.18532, \dodoi{10.48550/arXiv.2504.18532}

\bibitem[{Migdalek(2016)}]{M:16}
Migdalek, J. 2016, Journal of Physics B: Atomic, Molecular and Optical Physics, 49, 185004, \dodoi{10.1088/0953-4075/49/18/185004}

\bibitem[{{Molero} {et~al.}(2024){Molero}, {Matteucci}, {Spitoni}, {Rojas-Arriagada}, \& {Rich}}]{Molero2024}
{Molero}, M., {Matteucci}, F., {Spitoni}, E., {Rojas-Arriagada}, A., \& {Rich}, R.~M. 2024, \aap, 687, A268, \dodoi{10.1051/0004-6361/202450418}

\bibitem[{{Molero} {et~al.}(2021){Molero}, {Romano}, {Reichert}, {Matteucci}, {Arcones}, {Cescutti}, {Simonetti}, {Hansen}, \& {Lanfranchi}}]{Molero2021}
{Molero}, M., {Romano}, D., {Reichert}, M., {et~al.} 2021, \mnras, 505, 2913, \dodoi{10.1093/mnras/stab1429}

\bibitem[{{Molero} {et~al.}(2025){Molero}, {Magrini}, {Palla}, {Cescutti}, {Viscasillas V{\'a}zquez}, {Casali}, {Spitoni}, {Matteucci}, \& {Randich}}]{Molero2025}
{Molero}, M., {Magrini}, L., {Palla}, M., {et~al.} 2025, \aap, 694, A274, \dodoi{10.1051/0004-6361/202453466}

\bibitem[{{Montelius} {et~al.}(2022){Montelius}, {Forsberg}, {Ryde}, {J{\"o}nsson}, {Af{\c{s}}ar}, {Johansen}, {Kaplan}, {Kim}, {Mace}, {Sneden}, \& {Thorsbro}}]{montelius:22}
{Montelius}, M., {Forsberg}, R., {Ryde}, N., {et~al.} 2022, \aap, 665, A135, \dodoi{10.1051/0004-6361/202243140}

\bibitem[{{Moon} {et~al.}(2012){Moon}, {Wang}, {Park}, {Yuk}, {Chun}, \& {Jaffe}}]{Moon:2012}
{Moon}, B., {Wang}, W., {Park}, C., {et~al.} 2012, in Society of Photo-Optical Instrumentation Engineers (SPIE) Conference Series, Vol. 8450, Modern Technologies in Space- and Ground-based Telescopes and Instrumentation II, ed. R.~{Navarro}, C.~R. {Cunningham}, \& E.~{Prieto}, 845048, \dodoi{10.1117/12.925702}

\bibitem[{{Mura-Guzm{\'a}n} {et~al.}(2020){Mura-Guzm{\'a}n}, {Yong}, {Abate}, {Karakas}, {Kobayashi}, {Oh}, {Chun}, \& {Mace}}]{aldo:20}
{Mura-Guzm{\'a}n}, A., {Yong}, D., {Abate}, C., {et~al.} 2020, \mnras, 498, 3549, \dodoi{10.1093/mnras/staa2610}

\bibitem[{{Mura-Guzm{\'a}n} {et~al.}(2025){Mura-Guzm{\'a}n}, {Yong}, {Kobayashi}, {Tominaga}, {McKenzie}, {Salinas}, {Mace}, {Kim}, \& {Zucker}}]{aldo:25}
{Mura-Guzm{\'a}n}, A., {Yong}, D., {Kobayashi}, C., {et~al.} 2025, \mnras, 538, 3177, \dodoi{10.1093/mnras/staf464}

\bibitem[{{Nandakumar} {et~al.}(2023{\natexlab{a}}){Nandakumar}, {Ryde}, {Casagrande}, \& {Mace}}]{Nandakumar:2023}
{Nandakumar}, G., {Ryde}, N., {Casagrande}, L., \& {Mace}, G. 2023{\natexlab{a}}, \aap, 675, A23, \dodoi{10.1051/0004-6361/202346149}

\bibitem[{{Nandakumar} {et~al.}(2024{\natexlab{a}}){Nandakumar}, {Ryde}, {Forsberg}, {Montelius}, {Mace}, {J{\"o}nsson}, \& {Thorsbro}}]{Nandakumar:24_21elements}
{Nandakumar}, G., {Ryde}, N., {Forsberg}, R., {et~al.} 2024{\natexlab{a}}, \aap, 684, A15, \dodoi{10.1051/0004-6361/202348462}

\bibitem[{{Nandakumar} {et~al.}(2024{\natexlab{b}}){Nandakumar}, {Ryde}, {Hartman}, \& {Mace}}]{Nandakumar:ba}
{Nandakumar}, G., {Ryde}, N., {Hartman}, H., \& {Mace}, G. 2024{\natexlab{b}}, \aap, 690, A226, \dodoi{10.1051/0004-6361/202451255}

\bibitem[{{Nandakumar} {et~al.}(2023{\natexlab{b}}){Nandakumar}, {Ryde}, \& {Mace}}]{Nandakumar:2023b}
{Nandakumar}, G., {Ryde}, N., \& {Mace}, G. 2023{\natexlab{b}}, \aap, 676, A79, \dodoi{10.1051/0004-6361/202346875}

\bibitem[{{Nandakumar} {et~al.}(2022){Nandakumar}, {Ryde}, {Montelius}, {Thorsbro}, {J{\"o}nsson}, \& {Mace}}]{nandakumar:22}
{Nandakumar}, G., {Ryde}, N., {Montelius}, M., {et~al.} 2022, \aap, 668, A88, \dodoi{10.1051/0004-6361/202244724}

\bibitem[{{Nandakumar} {et~al.}(2025){Nandakumar}, {Ryde}, {Schultheis}, {Rich}, {di Matteo}, {Thorsbro}, \& {Mace}}]{NSC_all:25}
{Nandakumar}, G., {Ryde}, N., {Schultheis}, M., {et~al.} 2025, \apjl, 982, L14, \dodoi{10.3847/2041-8213/adbb6d}

\bibitem[{{Nave} {et~al.}(1994){Nave}, {Johansson}, {Learner}, {Thorne}, \& {Brault}}]{1994Nave}
{Nave}, G., {Johansson}, S., {Learner}, R.~C.~M., {Thorne}, A.~P., \& {Brault}, J.~W. 1994, \apjs, 94, 221, \dodoi{10.1086/192079}

\bibitem[{{Nishimura} {et~al.}(2017){Nishimura}, {Sawai}, {Takiwaki}, {Yamada}, \& {Thielemann}}]{Nishimura2017}
{Nishimura}, N., {Sawai}, H., {Takiwaki}, T., {Yamada}, S., \& {Thielemann}, F.~K. 2017, \apjl, 836, L21, \dodoi{10.3847/2041-8213/aa5dee}

\bibitem[{{Oh} {et~al.}(2024){Oh}, {Park}, {Kim}, {Kim}, {Jeong}, {Lee}, {Park}, {Yu}, {Kim}, {Chun}, {Oh}, {Jang}, {Jang}, {Seong}, {Lee}, {Kim}, {Lee}, {Ramos}, {Prado}, {Chinn}, {Arriagada}, {Diaz}, {White}, {Tapia}, {Xu}, {Suh}, {Miller}, {Stecher}, {Kurz}, {Quiroz}, {Figura}, {Hartman}, {Mocnik}, {Rawlings}, {Farina}, {Miller}, {Stephens}, {Oyarz{\'u}n}, {Olivares}, {Labrie}, {Hirst}, {Hayward}, {Brooks}, {Mace}, {Lee}, {Good}, {Jaffe}, {Kim}, {Yuk}, {Hwang}, \& {Park}}]{IGRINS2}
{Oh}, H., {Park}, C., {Kim}, S., {et~al.} 2024, in Society of Photo-Optical Instrumentation Engineers (SPIE) Conference Series, Vol. 13096, Ground-based and Airborne Instrumentation for Astronomy X, ed. J.~J. {Bryant}, K.~{Motohara}, \& J.~R.~D. {Vernet}, 1309608, \dodoi{10.1117/12.3017710}

\bibitem[{{Oliva} {et~al.}(2006){Oliva}, {Origlia}, {Baffa}, {Biliotti}, {Bruno}, {D'Amato}, {Del Vecchio}, {Falcini}, {Gennari}, {Ghinassi}, {Giani}, {Gonzalez}, {Leone}, {Lolli}, {Lodi}, {Maiolino}, {Mannucci}, {Marcucci}, {Mochi}, {Montegriffo}, {Rossetti}, {Scuderi}, \& {Sozzi}}]{giano:06}
{Oliva}, E., {Origlia}, L., {Baffa}, C., {et~al.} 2006, in Society of Photo-Optical Instrumentation Engineers (SPIE) Conference Series, Vol. 6269, Ground-based and Airborne Instrumentation for Astronomy, ed. I.~S. {McLean} \& M.~{Iye}, 626919, \dodoi{10.1117/12.670006}

\bibitem[{{Origlia} {et~al.}(2014){Origlia}, {Oliva}, {Baffa}, {Falcini}, {Giani}, {Massi}, {Montegriffo}, {Sanna}, {Scuderi}, {Sozzi}, {Tozzi}, {Carleo}, {Gratton}, {Ghinassi}, \& {Lodi}}]{Origlia:2014}
{Origlia}, L., {Oliva}, E., {Baffa}, C., {et~al.} 2014, in Society of Photo-Optical Instrumentation Engineers (SPIE) Conference Series, Vol. 9147, Ground-based and Airborne Instrumentation for Astronomy V, ed. S.~K. {Ramsay}, I.~S. {McLean}, \& H.~{Takami}, 91471E, \dodoi{10.1117/12.2054743}

\bibitem[{{Park} {et~al.}(2014){Park}, {Jaffe}, {Yuk}, {Chun}, {Pak}, {Kim}, {Pavel}, {Lee}, {Oh}, {Jeong}, {Sim}, {Lee}, {Nguyen Le}, {Strubhar}, {Gully-Santiago}, {Oh}, {Cha}, {Moon}, {Park}, {Brooks}, {Ko}, {Han}, {Nah}, {Hill}, {Lee}, {Barnes}, {Yu}, {Kaplan}, {Mace}, {Kim}, {Lee}, {Hwang}, \& {Park}}]{Park:2014}
{Park}, C., {Jaffe}, D.~T., {Yuk}, I.-S., {et~al.} 2014, in Society of Photo-Optical Instrumentation Engineers (SPIE) Conference Series, Vol. 9147, Ground-based and Airborne Instrumentation for Astronomy V, ed. S.~K. {Ramsay}, I.~S. {McLean}, \& H.~{Takami}, 91471D, \dodoi{10.1117/12.2056431}

\bibitem[{Pasquini \& Hubin(2018)}]{pasquini2018eso-485}
Pasquini, L., \& Hubin, N. 2018, in  (Ground-based and Airborne Instrumentation for Astronomy {VII}), 3 -- 11, \dodoi{10.1117/12.2313075}

\bibitem[{Piskunov {et~al.}(1995)Piskunov, Kupka, Ryabchikova, Weiss, \& Jeffery}]{vald}
Piskunov, N.~E., Kupka, F., Ryabchikova, T.~A., Weiss, W.~W., \& Jeffery, C.~S. 1995, A{\textbackslash} \&{AS}, 112, 525

\bibitem[{{Prantzos} {et~al.}(2020){Prantzos}, {Abia}, {Cristallo}, {Limongi}, \& {Chieffi}}]{prantzos:20}
{Prantzos}, N., {Abia}, C., {Cristallo}, S., {Limongi}, M., \& {Chieffi}, A. 2020, \mnras, 491, 1832, \dodoi{10.1093/mnras/stz3154}

\bibitem[{{Prantzos} {et~al.}(2018){Prantzos}, {Abia}, {Limongi}, {Chieffi}, \& {Cristallo}}]{prantzos:18}
{Prantzos}, N., {Abia}, C., {Limongi}, M., {Chieffi}, A., \& {Cristallo}, S. 2018, \mnras, 476, 3432, \dodoi{10.1093/mnras/sty316}

\bibitem[{{Rizzuti} {et~al.}(2019){Rizzuti}, {Cescutti}, {Matteucci}, {Chieffi}, {Hirschi}, \& {Limongi}}]{Rizzuti2019}
{Rizzuti}, F., {Cescutti}, G., {Matteucci}, F., {et~al.} 2019, \mnras, 489, 5244, \dodoi{10.1093/mnras/stz2505}

\bibitem[{{Romano} {et~al.}(2019){Romano}, {Matteucci}, {Zhang}, {Ivison}, \& {Ventura}}]{Romano2019}
{Romano}, D., {Matteucci}, F., {Zhang}, Z.-Y., {Ivison}, R.~J., \& {Ventura}, P. 2019, \mnras, 490, 2838, \dodoi{10.1093/mnras/stz2741}

\bibitem[{{Ryabchikova} {et~al.}(1997){Ryabchikova}, {Piskunov}, {Kupka}, \& {Weiss}}]{vald3}
{Ryabchikova}, T.~A., {Piskunov}, N.~E., {Kupka}, F., \& {Weiss}, W.~W. 1997, Baltic Astronomy, 6, 244

\bibitem[{{Ryde} {et~al.}(2020){Ryde}, {J{\"o}nsson}, {Mace}, {Cunha}, {Spitoni}, {Af{\c{s}}ar}, {Jaffe}, {Forsberg}, {Kaplan}, {Kidder}, {Lee}, {Oh}, {Smith}, {Sneden}, {Sokal}, {Strickland}, \& {Thorsbro}}]{Ryde:2020}
{Ryde}, N., {J{\"o}nsson}, H., {Mace}, G., {et~al.} 2020, \apj, 893, 37, \dodoi{10.3847/1538-4357/ab7eb1}

\bibitem[{{Ryde} {et~al.}(2025){Ryde}, {Nandakumar}, {Schultheis}, {Kordopatis}, {di Matteo}, {Haywood}, {Sch{\"o}del}, {Nogueras-Lara}, {Rich}, {Thorsbro}, {Mace}, {Agertz}, {Amarsi}, {Kocher}, {Molero}, {Orglia}, {Pagnini}, \& {Spitoni}}]{ryde:25}
{Ryde}, N., {Nandakumar}, G., {Schultheis}, M., {et~al.} 2025, \apj, 979, 174, \dodoi{10.3847/1538-4357/ad9b2b}

\bibitem[{{Sansonetti}(2006)}]{sansonetti:06}
{Sansonetti}, J.~E. 2006, Journal of Physical and Chemical Reference Data, 35, 301, \dodoi{10.1063/1.2035727}

\bibitem[{{Sawczynec} {et~al.}(2022){Sawczynec}, {Mace}, {Gully-Santiago}, \& {Jaffe}}]{rrisa}
{Sawczynec}, E., {Mace}, G., {Gully-Santiago}, M., \& {Jaffe}, D. 2022, in American Astronomical Society Meeting Abstracts, Vol.~54, American Astronomical Society Meeting Abstracts, 203.06

\bibitem[{{Sawczynec} {et~al.}(2025){Sawczynec}, {Kaplan}, {Mace}, {Lee}, {Jaffe}, {Park}, {Yuk}, {Chun}, {Pak}, {Hwang}, {Jeong}, {Kim}, {Kim}, {Kim}, {Kim}, {Le}, {Lee}, {Lee}, {Oh}, {Oh}, {Park}, {Park}, \& {Yu}}]{rrisa:25}
{Sawczynec}, E., {Kaplan}, K.~F., {Mace}, G.~N., {et~al.} 2025, \pasp, 137, 034505, \dodoi{10.1088/1538-3873/adba89}

\bibitem[{{Simonetti} {et~al.}(2019){Simonetti}, {Matteucci}, {Greggio}, \& {Cescutti}}]{Simonetti2019}
{Simonetti}, P., {Matteucci}, F., {Greggio}, L., \& {Cescutti}, G. 2019, \mnras, 486, 2896, \dodoi{10.1093/mnras/stz991}

\bibitem[{{Smith} {et~al.}(2021){Smith}, {Bizyaev}, {Cunha}, {Shetrone}, {Souto}, {Allende Prieto}, {Masseron}, {M{\'e}sz{\'a}ros}, {J{\"o}nsson}, {Hasselquist}, {Osorio}, {Garc{\'\i}a-Hern{\'a}ndez}, {Plez}, {Beaton}, {Holtzman}, {Majewski}, {Stringfellow}, \& {Sobeck}}]{Smith:2021}
{Smith}, V.~V., {Bizyaev}, D., {Cunha}, K., {et~al.} 2021, \aj, 161, 254, \dodoi{10.3847/1538-3881/abefdc}

\bibitem[{{Sneden} {et~al.}(2008){Sneden}, {Cowan}, \& {Gallino}}]{sneden:08}
{Sneden}, C., {Cowan}, J.~J., \& {Gallino}, R. 2008, \araa, 46, 241, \dodoi{10.1146/annurev.astro.46.060407.145207}

\bibitem[{{Spitoni} {et~al.}(2019){Spitoni}, {Silva Aguirre}, {Matteucci}, {Calura}, \& {Grisoni}}]{Spitoni2019}
{Spitoni}, E., {Silva Aguirre}, V., {Matteucci}, F., {Calura}, F., \& {Grisoni}, V. 2019, \aap, 623, A60, \dodoi{10.1051/0004-6361/201834188}

\bibitem[{{Takeda}(2021)}]{takeda:21}
{Takeda}, Y. 2021, Astronomische Nachrichten, 342, 515, \dodoi{10.1002/asna.202123873}

\bibitem[{{Taylor}(2005)}]{topcat}
{Taylor}, M.~B. 2005, in Astronomical Society of the Pacific Conference Series, Vol. 347, Astronomical Data Analysis Software and Systems XIV, ed. P.~{Shopbell}, M.~{Britton}, \& R.~{Ebert}, 29

\bibitem[{{Thorne} {et~al.}(1999){Thorne}, {Litzen}, \& {Johansson}}]{spectrophysics}
{Thorne}, A.~P., {Litzen}, U., \& {Johansson}, S. 1999, {Spectrophysics : principles and applications} ({Springer})

\bibitem[{{Tody}(1993)}]{IRAF}
{Tody}, D. 1993, in ASP Conf. Ser. 52: Astronomical Data Analysis Software and Systems II, ed. R.~J. {Hanisch}, R.~J.~V. {Brissenden}, \& J.~{Barnes}, 173

\bibitem[{{Tomkin} \& {Lambert}(1999)}]{tomkin:99}
{Tomkin}, J., \& {Lambert}, D.~L. 1999, \apj, 523, 234, \dodoi{10.1086/307735}

\bibitem[{{Valenti} \& {Piskunov}(1996)}]{sme}
{Valenti}, J.~A., \& {Piskunov}, N. 1996, \aaps, 118, 595

\bibitem[{{Valenti} \& {Piskunov}(2012)}]{sme_code}
---. 2012, {SME: Spectroscopy Made Easy}.
\newblock \doeprint{1202.013}

\bibitem[{van~der Walt {et~al.}(2011)van~der Walt, Colbert, \& Varoquaux}]{numpy}
van~der Walt, S., Colbert, S.~C., \& Varoquaux, G. 2011, Computing in Science and Engineering, 13, 22, \dodoi{10.1109/MCSE.2011.37}

\bibitem[{{Virtanen} {et~al.}(2020){Virtanen}, {Gommers}, {Oliphant}, {Haberland}, {Reddy}, {Cournapeau}, {Burovski}, {Peterson}, {Weckesser}, {Bright}, {van der Walt}, {Brett}, {Wilson}, {Millman}, {Mayorov}, {Nelson}, {Jones}, {Kern}, {Larson}, {Carey}, {Polat}, {Feng}, {Moore}, {VanderPlas}, {Laxalde}, {Perktold}, {Cimrman}, {Henriksen}, {Quintero}, {Harris}, {Archibald}, {Ribeiro}, {Pedregosa}, {van Mulbregt}, \& {SciPy 1. 0 Contributors}}]{scipy}
{Virtanen}, P., {Gommers}, R., {Oliphant}, T.~E., {et~al.} 2020, Nature Methods, 17, 261, \dodoi{10.1038/s41592-019-0686-2}

\bibitem[{{Wang} {et~al.}(2010){Wang}, {Gully-Santiago}, {Deen}, {Mar}, \& {Jaffe}}]{Wang:2010}
{Wang}, W., {Gully-Santiago}, M., {Deen}, C., {Mar}, D.~J., \& {Jaffe}, D.~T. 2010, in Society of Photo-Optical Instrumentation Engineers (SPIE) Conference Series, Vol. 7739, Modern Technologies in Space- and Ground-based Telescopes and Instrumentation, ed. E.~{Atad-Ettedgui} \& D.~{Lemke}, 77394L, \dodoi{10.1117/12.857164}

\bibitem[{{Warner}(1968)}]{1968MNRAS.139..115W}
{Warner}, B. 1968, \mnras, 139, 115, \dodoi{10.1093/mnras/139.1.115}

\bibitem[{{Wehrhahn} {et~al.}(2023){Wehrhahn}, {Piskunov}, \& {Ryabchikova}}]{pysme:23}
{Wehrhahn}, A., {Piskunov}, N., \& {Ryabchikova}, T. 2023, \aap, 671, A171, \dodoi{10.1051/0004-6361/202244482}

\bibitem[{{Wildi} {et~al.}(2017){Wildi}, {Blind}, {Reshetov}, {Hernandez}, {Genolet}, {Conod}, {Sordet}, {Segovilla}, {Rasilla}, {Brousseau}, {Thibault}, {Delabre}, {Bandy}, {Sarajlic}, {Cabral}, {Bovay}, {Vall{\'e}e}, {Bouchy}, {Doyon}, {Artigau}, {Pepe}, {Hagelberg}, {Melo}, {Delfosse}, {Figueira}, {Santos}, {Gonz{\'a}lez Hern{\'a}ndez}, {de Medeiros}, {Rebolo}, {Broeg}, {Benz}, {Boisse}, {Malo}, {K{\"a}ufl}, \& {Saddlemyer}}]{nirps}
{Wildi}, F., {Blind}, N., {Reshetov}, V., {et~al.} 2017, in Society of Photo-Optical Instrumentation Engineers (SPIE) Conference Series, Vol. 10400, Society of Photo-Optical Instrumentation Engineers (SPIE) Conference Series, ed. S.~{Shaklan}, 1040018, \dodoi{10.1117/12.2275660}

\bibitem[{{Yuk} {et~al.}(2010){Yuk}, {Jaffe}, {Barnes}, {Chun}, {Park}, {Lee}, {Lee}, {Wang}, {Park}, {Pak}, {Strubhar}, {Deen}, {Oh}, {Seo}, {Pyo}, {Park}, {Lacy}, {Goertz}, {Rand}, \& {Gully-Santiago}}]{Yuk:2010}
{Yuk}, I.-S., {Jaffe}, D.~T., {Barnes}, S., {et~al.} 2010, in Society of Photo-Optical Instrumentation Engineers (SPIE) Conference Series, Vol. 7735, Ground-based and Airborne Instrumentation for Astronomy III, ed. I.~S. {McLean}, S.~K. {Ramsay}, \& H.~{Takami}, 77351M, \dodoi{10.1117/12.856864}

\bibitem[{Özdemir {et~al.}(2025)Özdemir, Afşar, Sneden, VandenBerg, Denissenkov, Milone, Bozkurt, Oh, Sokal, Mace, \& Jaffe}]{odzemir:25}
Özdemir, S., Afşar, M., Sneden, C., {et~al.} 2025

\end{thebibliography}
\bibliographystyle{aasjournal}







\end{document}